\documentclass[notitlepage, nofootinbib, preprintnumbers, superscriptaddress, aps, reprint, prd]{revtex4-1}
\usepackage[utf8]{inputenc}
\usepackage{graphicx}
\usepackage{ulem}
\usepackage{hyperref}
\hypersetup{
	colorlinks   = true, 
	urlcolor     = blue, 
	linkcolor    = black, 
	citecolor   = red 
}
\usepackage{amsmath}
\usepackage{pgf}
\usepackage{amsfonts,mathtools}

\newcommand{\kv}{\textbf{k}}
\newcommand{\tpc}{(2 \pi)^3}

\newcommand{\taunl}{\tau_{\rm NL}}

\newcommand{\Bl}{\Big \langle}
\newcommand{\Br}{\Big \rangle}
\newcommand{\Aa}{\Delta\hat{X}_0}
\newcommand{\Ab}{\Delta\hat{X}_1}

\newcommand{\dv}{\textbf{d}}
\newcommand{\cov}{\textbf{C}}
\newcommand{\lnL}{\ln{\mathcal{L}}}
\newcommand{\Nextra}{\mathcal{N}}

\begin{document}
\preprint{IGC-18/4-1}
\title{Statistical anisotropies in temperature and polarization fluctuations from a scale-dependent trispectrum}
\author{Saroj Adhikari} \email{saroj@umich.edu}
\affiliation{Department of Physics, University of Michigan, 450 Church St, Ann Arbor, MI 48109-1040, U.S.A.}
\author{Anne-Sylvie Deutsch} \email{asdeutsch@psu.edu} \affiliation{Institute for Gravitation and the Cosmos, The Pennsylvania State University,\\ University Park Pennsylvania 16802}
\author{Sarah Shandera} \email{ses47@psu.edu} \affiliation{Institute for Gravitation and the Cosmos, The Pennsylvania State University,\\ University Park Pennsylvania 16802}

\date{\today}

\begin{abstract}
We study statistical anisotropies generated in the observed 
two-point function of the cosmic microwave background (CMB) fluctuations if the primordial statistics are non-Gaussian. Focusing on the dipole modulations of the anisotropies, we find that the hemispherical power asymmetry observed in the CMB {temperature fluctuations} can be modeled by a local-type trispectrum with amplitude $\taunl(k_p=0.05~{\rm  Mpc}^{-1}) \approx 2 \times 10^4$ and a large red tilt $n\approx -0.68$. We numerically evaluate the non-Gaussian covariance of the modulation estimators for both temperature and E-mode polarization fluctuations and discuss the prospects of constraining the model using {\it Planck} satellite data. We then discuss other effects of the scale-dependent trispectrum that could be used to distinguish this scenario from other explanations of the power asymmetry: higher-order modulations of the two-point function and the non-Gaussian angular power spectrum covariance. As an important consequence of the non-Gaussian power spectrum covariance, we discuss how the CMB-inferred spectral index of primordial scalar fluctuations can be significantly biased in the presence of a scale-dependent local-type trispectrum. 
\end{abstract}

\maketitle

\section{Introduction}
Several anomalies have been observed in the cosmic microwave background temperature fluctuations at the largest scales~\cite{Schwarz:2015cma}. They are measured features that are primordial and not due to instrumental noise or systematics, or due to late-time physics. Such anomalies are moderately unlikely to arise as mere statistical fluctuations in the Gaussian, isotropic models of cosmological fluctuations which otherwise describe observations with great precision. While the statistical significance of these unexpected features is not very strong, their presence has led to several model building attempts~\cite{Erickcek:2008sm, Erickcek:2009at, Lyth:2013vha, Liddle:2013czu, Byrnes:2016uqw, Dai:2013kfa, Ringeval:2015ywa, Kanno:2013ohv, Namjoo:2014nra, Kenton:2015jga, Byrnes:2015asa} aiming to constrain physics of the primordial universe.

The current ambiguous status of the anomalies on large scales \cite{PlanckIsotropy2015}, and in particular the hemispherical power asymmetry \cite{Eriksen:2007pc, Akrami:2014eta, Adhikari:2014mua} which has motivated this work, is driven by the fact that the large-scale temperature fluctuations have been measured to cosmic-variance limit. But, there is additional data available in principle, in particular from large-scale polarization \cite{Contreras2017}, from galaxy surveys \cite{Zhai:2017ibd}, from the scattering of CMB photons by free electrons after reionization~\cite{Terrana2016,Deutsch2017a,Deutsch2017}, or from 21-cm fluctuations \cite{Shiraishi:2016omb}.

The different efforts to model the observed statistical anisotropies in the CMB can be roughly categorized into two groups in which: (i) there is an explicit breaking of statistical isotropy \cite{Erickcek:2009at, Jazayeri:2014nya}, which means a preferred direction in the Universe, or (ii) the statistical isotropy breaking is spontaneous due to some stochastic modulating field \cite{Gordon:2005ai} or primordial non-Gaussianity \cite{Adhikari2016}. In this work, we use a framework where the observed power asymmetry arises spontaneously as the result of looking at a sub-volume of a larger space whose fluctuations are described by isotropic but non-Gaussian statistics. In a non-Gaussian model, the dipolar modulation of the Fourier space two-point function is described by the collapsed limit of the Fourier space four-point function (the trispectrum) of primordial fluctuations.

The relation between non-Gaussianity and statistical anisotropy has been discussed previously in the context of the CMB \cite{Ferreira:1997wd, Hanson2009, Dai:2013ikl}. In particular, {\it Planck} satellite data was used to constrain the amplitude $\tau_{\rm NL}$ of a scale-invariant local-type trispectrum by using statistical anisotropy estimators \cite{Hanson2009}, giving $\tau_{\rm NL}<2800$ at 95\% confidence level \cite{Ade:2013ydc}. (See also \cite{Feng:2015pva}.) However, since the observed asymmetry has a significant scale dependence, it is useful to expand on the {\it Planck} analysis and study in detail a scale-dependent trispectrum model. Non-Gaussianity that is scale dependent and larger on large scales can be consistent with the very tight scale-independent constraints, since those are driven by the many modes measurable on small scales. {An easy way to see this is to note that the $\taunl$ constraint from WMAP data, which are dominated by larger scales than {\it Planck} data, is an order of magnitude weaker \cite{Regan:2013jua}}.

We study the effect of a scale-dependent trispectrum in the CMB fluctuations by 
calculating the induced non-Gaussian covariance of modulation estimators. Such 
a formalism allows us to simultaneously consider the effect on the (correlated) 
modulations expected in CMB polarization and forecast the improvement in 
trispectrum constraints when adding polarization data. Further, we can also 
straightforwardly extend the study to other trispectra that have a large 
collapsed limit. The following primary results of our work all provide strong motivations to expand the current search for 
non-Gaussianities to the case of scale-dependent local-type trispectrum:
\begin{enumerate}
    \item Two of the large-scale CMB temperature anomalies --- the hemispherical power asymmetry and the power deficit at large scales --- can be well-modeled by a scale-dependent trispectrum, \\
    \item Such a trispectrum has other modulating effects on the temperature and polarization fluctuations that can be used to improve constraints on the scale-dependent trispectrum parameters, and \\
    \item If we require the trispectrum amplitude and parameters be large enough to explain both the hemispherical power asymmetry and the power deficit at large scales, then we find that the non-Gaussian covariance between the measured angular power spectra of the CMB can be  large enough to significantly bias the inference of cosmological parameters (see Figure \ref{fig:ngclcov}).
\end{enumerate} 

The rest of the paper is organized as follows. In Sec \ref{sec:anisotropy}, we discuss the general relationship between statistical anisotropies observed in a finite volume when the curvature fluctuations on larger scale are coupled to those on smaller scales. We then define modulation estimators in Sec \ref{sec:estimators} and describe how the effect of the non-Gaussian nature of fluctuations on the covariance of these estimators can be computed when a model for the primordial trispectrum is specified. In Sec \ref{sec:forecast}, we numerically evaluate these covariances and obtain a fiducial set of scale-dependent trispectrum parameters that can explain the observed hemispherical power asymmetry at large scales, and study how including polarization and higher-order modulations can improve model constraints. We discuss the non-Gaussian covariance of angular power spectra generated by a scale-dependent primordial trispectrum and how it can bias the reconstruction of the spectral index of the power spectrum in Sec \ref{sec:ngcovariance}. We summarize and conclude in Sec \ref{sec:summary}. 

\section{Spontaneous isotropy breaking from non-Gaussianity} \label{sec:anisotropy}
The statistics of the power asymmetry observed in a finite volume can be modeled as a spatial modulation of the observed temperature fluctuations. Simplifying to the scale-independent case for the moment, this is
\begin{equation}
\frac{\tilde{\delta T}}{T_0}=\frac{\delta T}{T_0}(1+A\hat{n}\cdot\hat{d}) 
\label{eq:Adipole}
\end{equation} 
where $\hat{n}$ is the direction of observation, $\hat{d}$ is the direction of the asymmetry, and $A$ is the amplitude. The standard Gaussian, isotropic model predicts that any given finite-sky realization will have an asymmetry drawn from a distribution with $\langle A\rangle=0$ and a finite variance determined by the power spectrum of the fluctuations. Models that introduce a new parameter for the asymmetry can either explicitly break isotropy, predicting a distribution with $\langle A\rangle\neq0$, or introduce a second, modulating Gaussian field that effectively boosts the variance of $A$ to be larger than expected from the measured isotropic fluctuations. For example, to boost the likelihood of an asymmetry only on large scales, one can introduce a field $h(x)$ into the primordial potential perturbations ($\Phi$), where $h(x)$ has fluctuations only on large scales and so is not a stochastic field in the finite volume \cite{Dvorkin2008}:
\begin{equation}
\label{eq:dvorkin}
\Phi(x)=g(x)[1+h(x)]
\end{equation}
In this case, $\langle g(x)h(x)\rangle=0$ and the observed asymmetry in a finite volume constrains the power spectrum of the second field. On the scales where the modulating field is stochastic, the curvature perturbations in this scenario have a connected four-point function proportional to the power in the two Gaussian fields ($P_g P_h$). For this reason, trispectrum estimators can be used, in the collapsed limit, to constrain the power asymmetry from such a modulating field.

In the simplest, scale-independent non-Gaussian scenario, the model for infinite volume statistics is
\begin{equation}
\Phi(x)=\sigma(x)+f_{\rm NL}[\sigma^2(x)-\langle\sigma^2(x)\rangle].
\label{eq:fnl}
\end{equation}
In any finite volume, the long wavelength modes of $\sigma$ play the role of the non-stochastic, modulating field exactly as in Eq.(\ref{eq:dvorkin}) previous case. The correlation between the power in short-wavelength modes and any gradient in the fluctuations of the long-wavelength modes results in spontaneous isotropy breaking observed in the finite volume. However, the extra variance of $A$ is not a new parameter to be constrained only by the asymmetry but is also constrained by isotropic non-Gaussianity in the finite volume (assuming the power spectrum is not suddenly very different on large scales). In the scenario in Eq.(\ref{eq:fnl}), there will be a Fourier space connected three-point function whose amplitude is proportional to $f_{\rm NL}$, and a four-point function with amplitude proportional to $f_{\rm NL}^2$. While the observed value of $f_{\rm NL}$ is itself subject to infrared divergent  cosmic variance in this case \cite{Nelson:2012sb,LoVerde:2013xka}, the non-Gaussian variance of $A$ is infra-red finite and is very nearly determined by the locally measured $f_{\rm NL}$ \cite{Adhikari2016}.

The models above are useful to contrast because they are simple\footnote{In this work, we use ``model" to refer to a description of the statistics of the inhomogeneities after reheating. We consider the construction of dynamical models of the primordial (inflationary) era that generates these statistics to be an extra layer, that likely provides an additional set of theoretical priors reflecting how difficult it seems to generate the respective dynamical models and how much inflation they tend to produce.}, but the class of scenarios of actual interest is more complicated since the observed asymmetry in CMB temperature is significantly scale dependent. Although constraints on non-Gaussianity have improved dramatically in recent years, there is still a surprising amount of space for non-Gaussian models to match existing data and generate the power asymmetry. The expression in Eq.(\ref{eq:fnl}) is an example of a much more general class of scenarios where the level of spontaneous isotropy breaking in a finite volume is enhanced by correlations between modes of different wavelengths. The momentum-space statistics of Eq.(\ref{eq:fnl}) have the feature that modes of very different wavelengths are coupled in a scale-independent way. A much wider range of couplings between modes of different wavelengths (including, of course, no coupling as is consistent with single-clock inflation) can be modeled by writing down particular 3-, 4-, ...$n$-point correlations in Fourier space and expanding the field $\Phi(\vec{k})$ in terms of a Gaussian field $\phi$ with appropriate kernels $K_n$ in the convolutions. Schematically \cite{Scoccimarro:2011pz, Baytas:2015nja}
\begin{equation}
\label{eq:NGfield}
\Phi(\vec{k})=\phi(\vec{k})+\left(\int \phi*\phi K_2\right)_{\vec{k}}+\left(\int \int\phi*\phi*\phi K_3\right)_{\vec{k}}+\dots
\end{equation}
Notice that there will be two contributions to the trispectrum: one that depends twice on the quadratic kernel $K_2$, and another from a single $K_3$ term. (In the local ansatz, these are the $\tau_{\rm NL}\propto f_{\rm NL}^2$ and $g_{\rm NL}$ terms, respectively.)

Even for a globally homogeneous, isotropic field, we expect some level of anisotropy if we restrict our observations to a sub-volume: for any single realization of the CMB sky, there is a direction that divides the map into two pieces with maximally different average amplitudes of power in the two halves. To see how the spontaneous breaking of isotropy is enhanced if the underlying field has a non-Gaussian component, we may divide the modes in Fourier space into long- and short-wavelength modes, and look at the expression for modes of the field that satisfy $k\gtrsim k_{\rm min}$:
\begin{align}
\Phi(\vec{k}_S)=&\phi(\vec{k}_S)\left[1+\int_{k_L} \phi K_2+\int_{k_L}\int_{k_L} \phi*\phi K_3+\dots\right]\nonumber \\ &+\left(\int \phi*\phi K_2\right)_{\vec{k}_S}\left[1+\int_{k_L}\phi K_3+\dots\right]\nonumber \\
&+\left(\int \int\phi*\phi*\phi K_3\right)_{\vec{k}_S}\left[1+\dots\right]+\dots
\label{eq:subVfield}
\end{align}
The fact that one may trade explicit isotropy breaking in a sub-volume for non-Gaussianity in an encompassing volume has been known for some time (see, eg the clear discussion in \cite{Hanson2009}). The calculation above, however, gives a straightforward means of generating both the isotropic and anisotropic statistics expected in a sub-volume for a wide range of models. {The statistical shift to the small-volume power spectrum (the linear term in Eq.(\ref{eq:subVfield})) can be expanded in spherical harmonics to give the expected level of isotropy breaking in the two-point function expected for a given model \cite{Adhikari2016}.}

From Eq.(\ref{eq:subVfield}), it is also clear that the  sub-volume statistics depend on parameters that control the size of all higher order, tree-level connected correlations. In general, then, the observed anisotropy is probing features of both the bispectrum and the trispectrum, and possibly beyond. An asymmetry in the observed power spectrum may be primarily generated by a four-point function if its amplitude (properly normalized by factors of the amplitude of fluctuations) is larger than that of the three-point function. {Particularly relevant for the observed power asymmetry is the case where the collapsed limit of the trispectrum is larger than it would be from the simple example of the single-source local ansatz given in Eq.(\ref{eq:fnl}). That is, the effective $\tau_{\rm NL}$ that governs the collapsed limit of the trispectrum is greater than $f_{\rm NL}^2$ \cite{Suyama:2007bg}.} In Eq.(\ref{eq:subVfield}), such an example requires a $K_4$ kernel that effectively subtracts off the $K_3^2$ contribution to the trispectrum and adds back the same shape, but with the appropriately scaled coefficient\footnote{An interesting test case of this type, that naturally gives a scale-dependent asymmetry, is that of quasi-single field inflation \cite{Chen:2009zp, Assassi:2012zq, Bonga:2015urq, Deutsch:2017rsn}. We find, however, that the quasi-single field parameters do not easily allow for $f_{\rm NL}$ small enough to be consistent with {\it Planck} constraints but $\tau_{\rm NL}$ as large as required by our fiducial model here (see Section \ref{sec:forecast}).}. An enhanced trispectrum, and power asymmetry, can also be generated by modifying Eq.(\ref{eq:NGfield}) to allow two separate fields additively sourcing the curvature (eg, Gaussian field + non-Gaussian field).  

\section{Modulations in the CMB fluctuations} \label{sec:estimators}
In this section, we describe and compute statistics of CMB modulations from a scale-dependent primordial trispectrum. The observed statistics in the CMB are the multipole moments of temperature or polarization fluctuations, which  depend on the primordial potential $\Phi(\kv)$ as follows
\begin{align}
    a_{\ell m}^x &= 4\pi (-i)^\ell \int \frac{d^3\kv}{\tpc} \Phi(\kv) g_\ell^x(k) Y_{\ell m}^*(\hat{k}),
\end{align}
where $g^x_\ell(k)$ is the CMB transfer function with $x=T,E$ describing temperature and E-mode polarization fluctuations respectively.

The role of polarization fluctuations in helping pin down whether the modulations observed in the temperature fluctuations are primordial or not has been previously studied in \cite{Dvorkin2008, Namjoo:2014pqa, Contreras2017}. Here we write down the general expressions for the covariances of modulation estimators in the presence of a trispectrum. We will use them to generate realizations of the estimators and study the expected constraints by using {\it Planck} temperature and polarization data in the next section. For the most part, we will focus on dipole modulations of the cosmic microwave background fluctuations (both T and E). 

Let us define the dipole modulation estimators using $\ell, \ell+1$ correlations as follows:
\begin{align}
\Delta \hat{X}_0^{wx}(\ell) &= \frac{1}{(2\ell+1) \sqrt{C_\ell^{ww} C_{\ell+1}^{xx}}} \sum_{m=-\ell}^\ell a^{w*}_{\ell m} a^x_{\ell+1,m} \\
\Delta \hat{X}_1^{wx}(\ell) &= \frac{1}{(2\ell+1) \sqrt{C_\ell^{ww} C_{\ell+1}^{xx}}} \sum_{m=-\ell}^{\ell} a_{\ell m}^{w*} a^x_{\ell+1, m+1} 
\label{eq:estimators}
\end{align}
where $w,x$ can be either $T,E$ and $C_\ell$s are the CMB angular power spectrum of the best-fit cosmology. (Note that, while we use the notation from \cite{Contreras2017} of $\Delta \hat{X}_M$s, our definition does not include additional $\ell$-dependent factors that exactly map the $\ell, \ell+1$ correlations to the Cartesian components of dipole modulation parameter $A$ as defined as in Eq.(\ref{eq:Adipole}).) Similar estimators can be defined for higher-order modulations, by considering $\ell, \ell+2$ correlations, for example for quadrupolar modulation. If the primordial fluctuations are Gaussian, the covariance of the dipole modulation estimators is given by
\begin{align}
\Bl \Delta \hat{X}^{wx*}_M(\ell) \Delta\hat{X}^{yz}_{M'}(\ell') \Br_{\rm G} =& \frac{\delta_{M,M'}\delta_{\ell,\ell'}}{2\ell+1} \nonumber \\ &\frac{C_\ell^{wy} C_{\ell+1}^{xz}}{\sqrt{C_\ell^{ww} C_{\ell+1}^{xx} C_{\ell'}^{yy} C_{\ell'+1}^{zz}}},
\end{align}
where $M,M'=0,1$. Note that $\Delta \hat{X}_0(\ell)$ are real, whereas $\Delta \hat{X}_1(\ell)$ are complex and the three degrees of freedom among the two estimators determine the amplitude and direction of the dipole modulation.

For models that generate a CMB power asymmetry by explicitly changing the power spectrum (for example, assuming that the primordial power spectrum has a dipole modulation), the means of $\Delta \hat{X}_M$ are non-zero: $\langle\Delta \hat{X}_M(\ell) \rangle \neq 0$. However, in models where the primordial fluctuations have significant non-Gaussianity, it is possible that global isotropy is respected, i.e. $\langle\Delta \hat{X}_M(\ell)\rangle = 0$, but the expected cosmic variance of CMB dipolar modulation increases. The resulting apparent statistical anisotropy is a spontaneous statistical isotropy breaking \cite{Gordon:2005ai} caused by the non-Gaussian nature of fluctuations. 
In that case, the non-Gaussian contribution to the covariance depends on a particular configuration of the CMB trispectrum: 
\begin{align}
\Bl \Delta &\hat{X}^{wx*}_M(\ell) \Delta\hat{X}^{yz}_{M'}(\ell') \Br_{\rm nG} \nonumber \\ 
&= \delta_{M, M'}\frac{\sum_{m, m'} \Bl a_{\ell m}^w\; a^{x*}_{\ell+1, m+M}\; a^{y*}_{\ell' m'}\; a_{\ell' +1, m'+M'}^z \Br_c}{(2\ell+1)(2\ell'+1)\sqrt{C_\ell^{ww} C_{\ell+1}^{xx} C_{\ell'}^{yy} C_{\ell'+1}^{zz}}}
\label{eq:DX0cov}
\end{align}
where the subscript $_c$ indicates connected part of the trispectrum.

To compute the CMB four-point function we follow the method in \cite{Hu:2001fa, 
Okamoto:2002ik}, which constructs the CMB trispectrum from a ``reduced 
trispectrum" that automatically enforces the trispectrum to have rotation, 
parity and permutation symmetries. The CMB four-point function can be written using Wigner-3j symbols, as:

\begin{align}
     \Bl a_{\ell_1 m_1}^w & a_{\ell_2 m_2}^x a_{\ell_3 m_3}^y a_{\ell_4 m_4}^z 
     \Br_c \nonumber \\  =& \sum_{LM} P^{w\ell_1 x\ell_2}_{y\ell_3 z\ell_4}(L) 
     \left(\begin{array}{ccc} \ell_1 & \ell_2 & L \\ m_1 & m_2 & -M \end{array} 
     \right) \nonumber \\ & \left(\begin{array}{ccc} \ell_3 & \ell_4 & L \\ m_3 
     & m_4 & M \end{array} \right)(-1)^M + (\ell_2 
     \leftrightarrow \ell_3) + (\ell_2 \leftrightarrow \ell_4)
    \label{eq:cmbfourpointfunction}
\end{align}
where \cite{Hu:2001fa}:
\begin{align}
    P^{w\ell_1 x\ell_2}_{y\ell_3 z\ell_4}(L) = \mathcal{T}^{w\ell_1 
    x\ell_2}_{y\ell_3 z\ell_4}(L) &+ (-1)^{L+\ell_1+\ell_2} 
    \mathcal{T}^{x\ell_2 
    w\ell_1}_{y\ell_3 z\ell_4}(L) \nonumber \\ &+ (-1)^{L+\ell_3+\ell_4}   
    \mathcal{T}^{w\ell_1 x\ell_2}_{z\ell_4 y\ell_3}(L) \nonumber \\  &+ 
    (-1)^{\ell_1+\ell_2+\ell_3+\ell_4} \mathcal{T}^{x\ell_2 w\ell_1}_{z\ell_4 
    y\ell_3}(L)
\label{eq:p}
\end{align}

The reduced CMB trispectrum $\mathcal{T}$ depends on the model of primordial trispectrum. In this work, we will consider a scale-dependent local $\taunl$ trispectrum \cite{Byrnes:2010ft}
\begin{align}
    T(\kv_1, \kv_2, \kv_3, \kv_4) &= \taunl \left(\frac{k_2 k_4}{k_p^2}\right)^n P(k_1) P(k_3) P(|\kv_1-\kv_2|) \nonumber \\ 
    &~~~ + {\rm permutations}
    \label{eq:trispectrum}
\end{align}
where the index $n$ describes the scale dependence of the trispectrum amplitude of the otherwise local-type trispectrum, and $k_p$ is the pivot at which $\taunl$ is the amplitude; we take $k_p=0.05 {\;\rm Mpc}^{-1}$. 
Similar to the calculation for the constant $\taunl$ trispectrum \cite{Okamoto:2002ik, Regan:2010cn}, we obtain
\begin{align}
    \mathcal{T}^{w\ell_1 x\ell_2}_{y\ell_3 z\ell_4}(L) = {\taunl} h_{\ell_1 
    \ell_2 L} h_{\ell_3 \ell_4 L} \int dr_1 r_1^2 \alpha_{\ell_1}^w(r_1, n) 
    \beta_{\ell_2}^x(r_1) \nonumber \\ \int dr_2 r_2^2 \alpha_{\ell_3}^y(r_2, 
    n) \beta_{\ell_4}^z(r_2) F_L(r_1, r_2)
    \label{eq:reducedtrispectrum}
\end{align}
where
\begin{align}
 h_{\ell_1 \ell_2, L} &= \sqrt{\frac{(2\ell_1+1)(2\ell_2+1)(2L+1)}{4\pi}} \left(\begin{array}{ccc} \ell_1 & \ell_2 & L \\ 0 & 0 & 0 \end{array}\right) \\
\alpha_\ell^w(r, n) &= \frac{2}{\pi} \int\; dk\;  k^2 \left(\frac{k}{k_p}\right)^{n}
g_\ell^w(k) j_\ell(kr) \\
\beta_\ell^x(r)&= 4\pi \int \frac{dk}{k} \mathcal{P}_\Phi(k) g_\ell^x(k) 
j_\ell(kr) \\
F_L(r_1, r_2) &= 4\pi \int \frac{dK}{K} \mathcal{P}_\Phi(K) j_L(K r_1) j_L(K r_2)
\end{align}
{and the $j_{\ell}$ are spherical Bessel functions. Note that the angular power spectrum can be written as an integral over the comoving distance $r$, using the quantities $\alpha_\ell(r)$ and $\beta_\ell(r)$ defined above:}
  \begin{align}
  C_\ell^{wx} &= \int \;dr\; r^2 \alpha_\ell^w(r, 0) \beta_\ell^x(r).
  \end{align}

Numerically evaluating the reduced trispectrum of Eq.(\ref{eq:reducedtrispectrum}) allows us to compute the non-Gaussian covariances, Eq.(\ref{eq:DX0cov}), for the dipole modulation estimators. The full covariance matrix for dipole modulation estimators (including the $f_{\rm sky}$ scaling for partial sky coverage and the noise power spectra) is given by,
\begin{widetext}
    \begin{align}
    \cov &= \Bl \Delta \hat{X}^{wx*}_M(\ell) \Delta\hat{X}^{yz}_{M'}(\ell') \Br \nonumber \\ &= \frac{1}{(2\ell+1)f_{\rm sky}} \frac{\delta_{M,M'}}{{\sqrt{C_\ell^{ww} C_{\ell+1}^{xx} C_{\ell'}^{yy} C_{\ell'+1}^{zz}}}} \left[ \delta_{\ell,\ell'} \tilde{C}_\ell^{wy} \tilde{C}_{\ell+1}^{xz} + \frac{1}{2\ell'+1} \sum_{m,m'} \Bl a_{\ell m}^w a^{x*}_{\ell+1, m+M} a^{y*}_{\ell' m'} a_{\ell'+1, m'+M'}^{z} \Br_c \right],
    \label{eq:dipolecov}
    \end{align}
\end{widetext}
where $w,x,y,z$ can be $T,E$, while $M,M'=0,1$ (of the $\Delta \hat{X}_{0,1}$) and $\tilde{C}_\ell^{wy} = C^{wy}_{\ell,\rm cmb} + C^{wy}_{\ell,\rm noise}$. The noise power spectrum for {\it Planck} is approximated using the specifications for two channels as in \cite{Galli:2014kla} with $f_{\rm sky}=0.65$. For numerical evaluations, we use \texttt{camb} \cite{Lewis:1999bs} to obtain the transfer functions $g_\ell(k)$ using {\it Planck} 2015 best-fit cosmological parameters \cite{Ade:2015xua}. We also follow the approximation outlined in Appendix for faster numerical evaluation, and mostly limit ourselves to $\ell \geq 30$ for which the approximation is correct within a few percent. {In the next section, we use realizations of $\Delta \hat{X}_M$s obtained using the full covariance matrix Eq.(\ref{eq:dipolecov}) to obtain our fiducial scale-dependent trispectrum parameters: $\taunl = 2\times 10^4, n=-0.68$.}

\begin{figure}
    \includegraphics[width=0.48\textwidth]{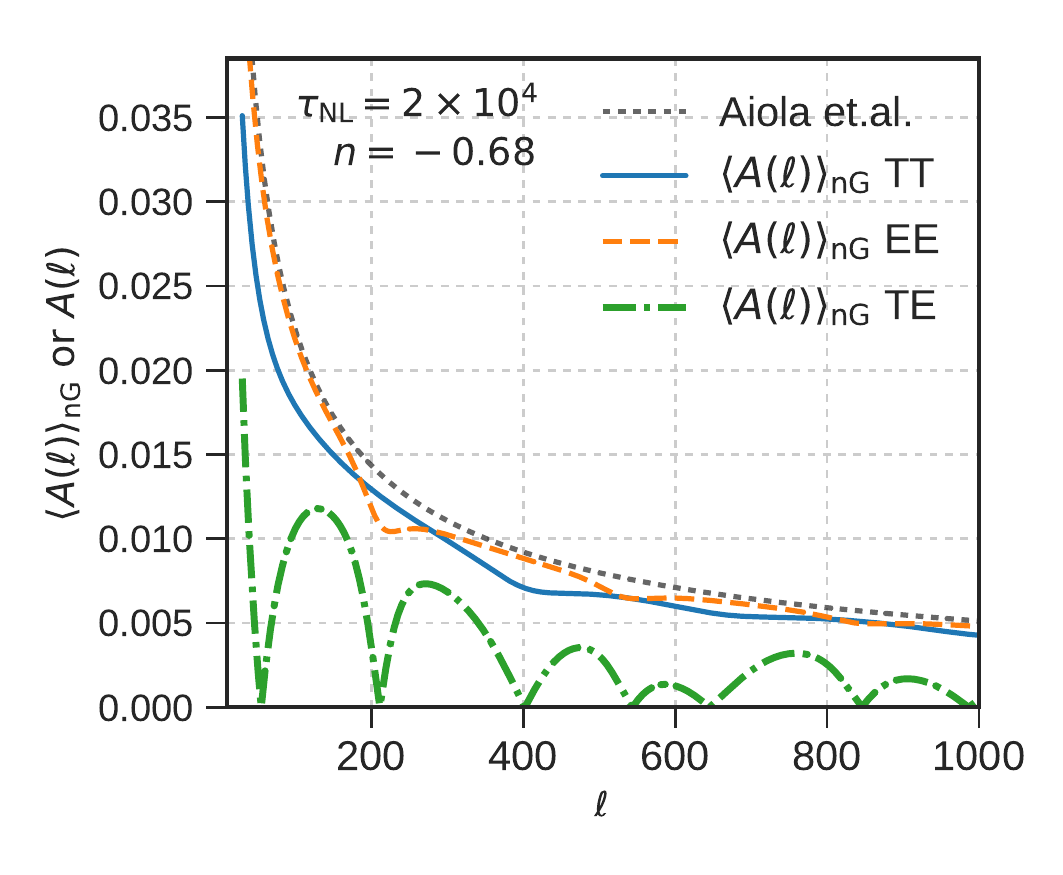}
    \caption{The expected $\ell$ 
    dependent amplitude of dipole modulation from the scale-dependent trispectrum model given in Eq.(\ref{eq:trispectrum}) for our fiducial parameters. For reference, we have plotted the best-fit $\ell-$dependent dipole modulation amplitude obtained by Aiola et. al \cite{Aiola:2015rqa} using Planck temperature data. The $\langle A(\ell)\rangle_{\rm nG}$ TT (solid blue) values only include the non-Gaussian contribution (and no Gaussian noise) which could explain its smaller magnitudes than that of the best-fit Aiola et.al. model (black dotted).}
    \label{fig:diagonalTE}
\end{figure}

In Figure \ref{fig:diagonalTE}, we plot the expectation value of the non-Gaussian contribution to the dipole modulation amplitude, 
\[ \Bl A(\ell)\Br_{\rm nG} \equiv \sqrt{\Bl A_x^2(\ell)\Br_{\rm nG} + \Bl A_y^2(\ell)\Br_{\rm nG} + \Bl A_z^2(\ell)\Br_{\rm nG}}\] 
using the fiducial $\taunl, n$ values. Here $A(\ell)$ corresponds to the harmonic transform of a dipolar modulation of the fluctuations $A(\hat{n}\cdot \hat{d})$ as in Eq.(\ref{eq:Adipole}). Note that there are additional $\ell$ dependent factors between our definition of $\Delta \hat{X}_M$s and the Cartesian components of $A$, which approach constant values at large $\ell$: $\Aa \approx (4/5) A_z, {\rm Re}\Ab \approx A_x/2, {\rm Im}\Ab \approx A_y/2$; we account for these factors between $\Delta \hat{X}_M$s and $A_{x,y,z}$s when computing $A(\ell)$ and comparing our results to that of \cite{Aiola:2015rqa}, which fitted the {\it Planck} temperature dipole modulation data to a phenomenological $\ell-$dependent model: 
    \begin{align}
        A(\ell) &= A_{\ell_0}\left(\frac{\ell}{\ell_0}\right)^n.
        \label{eq:Aiolamodel}
    \end{align}
From Figure \ref{fig:diagonalTE}, we can see that a scale-dependent trispectrum can generate a scale-dependent dipole modulation of the CMB temperature fluctuations similar to the best-fit values found by \cite{Aiola:2015rqa}. We have also plotted the corresponding scale-dependent dipole modulations expected in EE and TE spectra. 

In Figure \ref{fig:diagonalTE}, we see that in general the polarization asymmetry amplitude is larger than that of the temperature. See also \cite{Namjoo:2014pqa, Contreras2017} for similar results and discussion. The reason is that temperature multipoles get contribution from a wider range of scales, and each modulation multipole roughly traces the average level of modulation over this range of scales. The transfer functions for E-modes, however, are generally narrower in range of wavenumber $k$ and trace, on average, larger scales compared to the temperature fluctuations. The modulation amplitude in our fiducial model decreases at smaller scales, which results in larger E-mode modulation. If, however, we postulate a scale dependence of the modulation amplitude which increases at smaller scales (i.e. $n>0$), the temperature modulation amplitudes on average will be larger than the polarization modulation amplitudes.

\section{Forecast} \label{sec:forecast}
In this section, we use numerical evaluations of the non-Gaussian covariances described in the previous section and generate various realizations of modulation parameters --- at different multipoles --- for both temperature and polarization fluctuations. We make use of these realizations to choose fiducial trispectrum model parameters by selecting ``look-alike" realizations in temperature fluctuations over a range $\ell=30-600$, where there is $3\sigma$ evidence for hemispherical power asymmetry. The polarization realizations that are generated simultaneously with the temperature realizations are then used to forecast the prospect of detecting the fiducial trispectrum model by using the log-likelihood difference. 

\subsection{Studies on $\Delta \hat{X}_M(\ell)$ realizations and fiducial model}
\begin{figure}
    \includegraphics[width=0.48\textwidth]{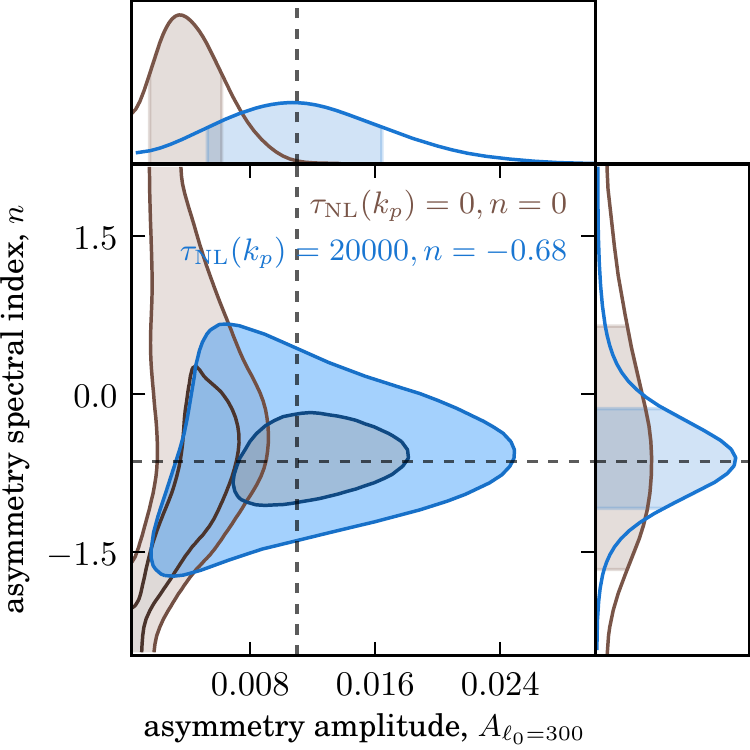}
    \caption{Distributions (smoothed histograms) of scale-dependent power asymmetry parameters $(A_{\ell_0=300},n)$ for two different models: (i) a Gaussian and isotropic model (brown, contour with smaller amplitudes), and (ii) our fiducial non-Gaussian isotropic model (blue, contour with larger amplitudes). For all the distributions, we have used temperature multipoles $\ell=30-600$, and assumed $f_{\rm sky}=0.65$. These smoothed distributions are generated from 25,000 modulation data realizations and the two contours indicate $1\sigma$ and $2\sigma$ intervals. The dashed lines show the marginalized median values obtained from fit to {\it Planck} data by \cite{Aiola:2015rqa}.}
    \label{fig:Anrandom}
\end{figure}

Using the full covariance matrix Eq.(\ref{eq:dipolecov}) we generate realizations of modulation parameters \[\dv = \{\Delta \hat{X}_0(\ell), {\rm Re}\Ab(\ell), {\rm Im} \Ab(\ell)\}\] for various values of $\taunl, n$. From the realizations, we measure the best-fit dipole amplitude and scale dependence $(A,n)$ by fitting to the function Eq.(\ref{eq:Aiolamodel}): $A(\ell) = A(\ell_0)(\ell/\ell_0)^n$; we choose $\ell_0=300$ instead of $\ell_0=60$ as in \cite{Aiola:2015rqa} but translate their constraints accordingly. Based on the distributions of $A,n$ obtained using realizations for a number of $(\taunl, n)$ values, we choose our fiducial scale-dependent trispectrum parameter values to be $\taunl = 2\times 10^4, n=-0.68$. The non-Gaussian model using these fiducial parameters produces median amplitude and scale dependence for the temperature modulation similar to the marginalized values found in \cite{Aiola:2015rqa}, indicated by the dashed lines in Figure \ref{fig:Anrandom}. We contrast the $(A,n)$ distribution generated by our fiducial trispectrum model to the distribution of $(A,n)$ obtained from Gaussian, isotropic realizations. The 2D smoothed histograms obtained by these two set of realizations are shown in Figure \ref{fig:Anrandom}. The shape of the contours indicates that even a significantly smaller $\taunl$ would make the observed power asymmetry more likely compared to $\taunl=0$.

Interestingly, the rough estimates for the scale-dependent trispectrum parameters in an inflationary model that generates dipole asymmetry while respecting current bispectrum constraints given in \cite{Byrnes:2015dub} are similar to our fiducial parameters. However, it is important to note that an actual data analysis of the temperature modulation data to fit for the scale-dependent trispectrum parameters hasn't been done yet, which may result in different values than our fiducial model --- especially when larger multipoles and higher-order modulations are included.

\begin{figure}
    \centering
    \includegraphics[width=0.48\textwidth]{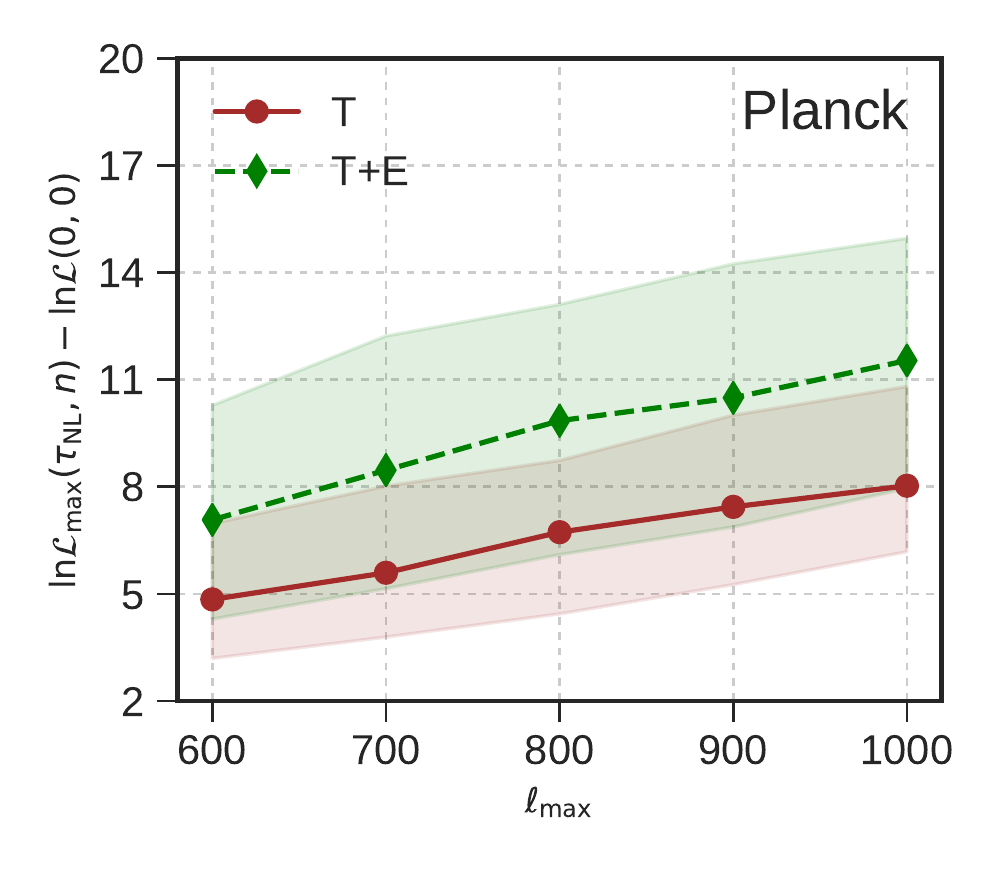}
    \caption{The log-likelihood improvement as a function of maximum multipole used in the analysis using {\it Planck}-like temperature $\Delta \hat{X}_M^T$ realizations only, and in combination with {\it Planck}-like E-mode $\Delta \hat{X}_M^E$ realizations. We find that, for our fiducial trispectrum model, adding E-mode polarization data from {\it Planck} increases the $\Delta \ln \mathcal{L} \sim 3$ over the temperature-only data. The data points in the figure are median $\Delta \ln \mathcal{L}$ values from fit to realizations while the band shows the $68\%$ spread around it.}
    \label{fig:lnLike}
\end{figure}

To examine the constraints on the scale-dependent parameters from {\it Planck} temperature data (and information added by polarization data), we select ``look-alike" realizations from our set of fiducial non-Gaussian realizations that produce scale-dependent dipole modulation in temperature fluctuations similar to the best fit values found in \cite{Aiola:2015rqa} (within ten percent of $A_{300}=0.011, n=-0.64$). We then compute the log-likelihood improvement with respect to the isotropic and Gaussian model, ($\Delta \ln \mathcal{L} \equiv \ln \mathcal{L}_{\rm max}(\taunl, n) - \ln \mathcal{L}(0,0)$), where
\begin{align}
\lnL(\taunl, n)= -\frac{1}{2} \left[{\rm det}\cov + \dv^T \cov^{-1} \dv\right]
\end{align}
and $\cov$ is a function of $(\taunl, n)$ given in Eq.(\ref{eq:dipolecov}). The estimated log-likelihood improvement $\Delta \ln \mathcal{L}$ as a function of the maximum multipole used to compute the log-likelihoods is plotted in Figure \ref{fig:lnLike}. Using only the temperature dipole modulation data upto $\ell_{\rm max}=1000$ we estimate a log-likelihood improvement of $\Delta \ln \mathcal{L} \sim 8$ for two extra parameters over the isotropic Gaussian model. We limit to $\ell_{\rm max}=1000$ for two reasons: (i) the signal for the models with strong scale dependence decreases sharply at larger multipoles, and (ii) the Doppler and aberration contribution starts to become important at higher multipoles that needs to be accounted \cite{Quartin:2014yaa}.

\subsection{Addition of polarization data}
To investigate the expected improvement by adding $E-$mode polarization data, we select the E-mode realizations corresponding to the ``look-alike" temperature realizations with power asymmetry similar to what is observed; recall that this is done assuming our fiducial model parameters: $\taunl(k_p) = 2\times 10^4, n=-0.68$. Also, the polarization realizations are generated simultaneously with the temperature realizations to account for both the Gaussian and non-Gaussian covariances between $T$ and $E$. Then, we quantify the contribution from the addition of polarization data by the improvement in log-likelihood, plotted in Figure \ref{fig:lnLike} by including TT, EE and TE modulation estimators.

The results are plotted in Figure \ref{fig:lnLike}. Even with the noise levels of {\it Planck}, we expect significant improvement in log-likelihood $\Delta\ln \mathcal{L} \sim 3$ by adding EE and TE dipole modulation data, for our fiducial trispectrum model.

\subsection{Expected improvement by using higher-order modulations}
We estimate the improvement when including higher-order ($L>1$) modulations by calculating the signal-to-noise ratio from the connected CMB trispectrum, for each modulation order, using \cite{Okamoto:2002ik, Kogo:2006kh}:
\begin{align}
\left(\frac{S}{N}\right)^2_L \approx& \sum_{\substack{\ell_1\geq\ell_2\\ \ell_2\geq\ell_3\\\ell_3\geq\ell_4}}^{\ell_{\rm max}} \sum_{abcd}\sum_{wxyz} {T}^{w_{\ell_1} x_{\ell_2}}_{y_{\ell_3} z_{\ell_4}}(L) \left[{\rm Cov}\right]^{-1} {T}^{a_{\ell_1} b_{\ell_2}}_{c_{\ell_3} d_{\ell_4}}(L), \\
\left[{\rm Cov}\right] =& (2L+1) C_{\ell_1}^{aw}C_{\ell_2}^{bx}C_{\ell_3}^{cy}C_{\ell_4}^{dz}
\end{align}
where $abcd,wxyz=\{{\rm TTTT, EEEE, TETE}\}$ for information using the TT, EE and TE modulation estimators (note that there are other combinations possible when using the trispectrum directly rather than using modulations), in which case the covariance [Cov] for each unique set of $\ell_1, \ell_2, \ell_3, \ell_4$ is a $3\times 3$ matrix. We find that for our fiducial model, \begin{align}
    \sum_L\left(\frac{S}{N}\right)^2_{L=2,3,4} = 0.3\left(\frac{S}{N}\right)^2_{L=1}
\end{align}
so, we can expect  $\sim 30\%$ increase in $\Delta \ln \mathcal{L}_{\rm max}$ shown in Figure \ref{fig:lnLike} by adding $L=2,3,4$ modulations of TT, EE, TE from {\it Planck}.

The use of higher-order modulations can help distinguish between a primordial 
trispectrum and a model in which the primordial power spectrum has a genuine 
statistical anisotropy. In the latter case, the preferred direction for 
higher-order modulations is the same as that of the dipole modulation. However, 
for a non-Gaussian model, the dipole and quadrupole modulation directions are 
uncorrelated as $\langle a_{\ell m}^w a_{\ell+1, m}^{x*} a_{\ell m}^{y*} 
a_{\ell+2, m}^z\rangle_c = 0$ for a parity-invariant primordial trispectrum.

\begin{figure*}
    \includegraphics[width=0.48\textwidth]{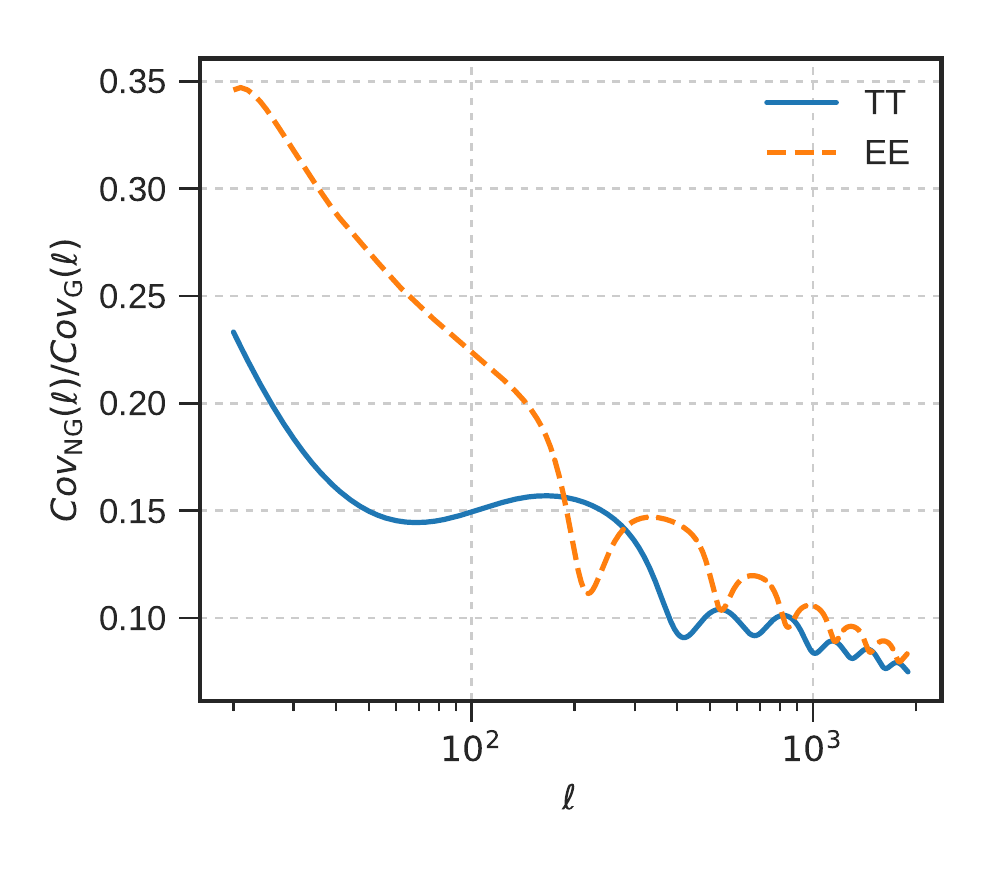}
    \includegraphics[width=0.48\textwidth]{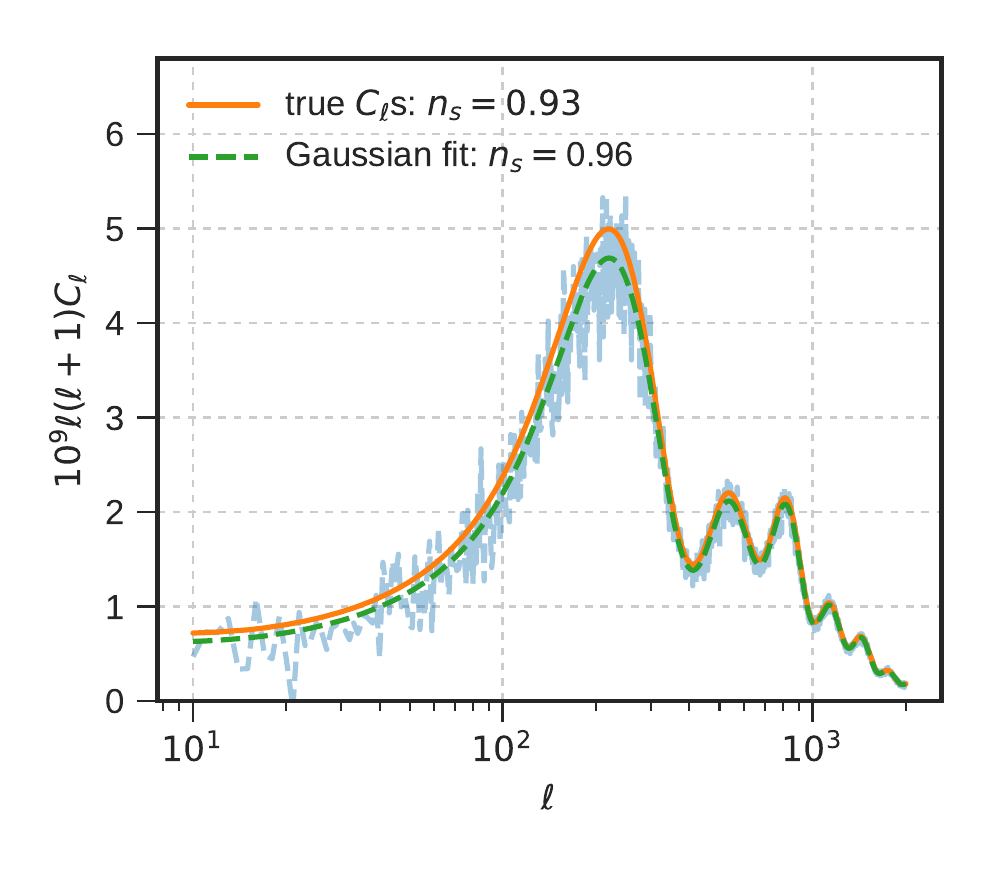}
    \caption{\textit{Left}: The diagonal component of the non-Gaussian term in the power spectrum covariance as a fraction of the Gaussian term for $(\taunl, n)=(2\times 10^4, -0.68)$ model with $\Nextra = 40$. \textit{Right}: Example of how the inferred spectral index $n_s$ can be significantly biased high if the observed large-scale power deficit is due to a scale-dependent trispectrum. In this example,  a set of $C_\ell$ realizations was generated using the non-Gaussian model with spectral index (at the pivot wavenumber $k_0=0.05 {\rm Mpc^{-1}}$) $n_s=0.93$, and the particular realization in the figure is a typical ($\sim 1\sigma$) realization of the non-Gaussian model with  $\sim 12\%$ power deficit for $\ell = 10-40$. The Gaussian fit value of spectral index was obtained by only varying $A_s, n_s$ and keeping other non-primordial parameters fixed.}
    \label{fig:ngclcov}
\end{figure*}

\section{Isotropic modulation of the power spectrum} \label{sec:ngcovariance}
A trispectrum with large collapsed-limit signal also modulates the {isotropic} angular power spectrum (the $C_\ell$) of the CMB. The collapsed-limit configuration of the trispectrum induces covariance between measured angular power spectra at widely separated multipoles. For the case of a constant modulation or a scale-invariant local trispectrum, the effect only rescales the amplitude of fluctuations. However, as we show below, for a scale-dependent trispectrum there can be more interesting effects. The covariance between measured angular power spectra in the presence of a non-zero connected trispectrum is given by,

\begin{align}
\cov(\hat{C}_\ell, \hat{C}_{\ell'}) =& \frac{2 C_\ell^2}{2\ell+1} \delta_{\ell, \ell'} + \frac{1}{(2\ell+1)(2\ell'+1)} \nonumber \\ &\sum_{m, m'} \Bl a_{\ell m} a_{\ell,-m} a_{\ell', m'} a_{\ell', -m'} \Br_c
\label{eq:Clcov}
\end{align}
where,
\begin{align}
\hat{C}_\ell = \frac{1}{2\ell+1} \sum_{m=-\ell}^\ell a_{\ell m}^* a_{\ell m}.
\end{align}

In standard cosmological analyses, the second term in Eq.(\ref{eq:Clcov}) is ignored assuming that the primordial fluctuations are Gaussian distributed. Inclusion of the second term, if non-zero and known (say from direct trispectrum measurements), can increase the measurement errors on cosmological parameters, which can be estimated through a Fisher analysis \cite{Hu:2001fa}. In the absence of tight scale-dependent trispectrum constraints, one can marginalize over the non-Gaussian covariance, which may however significantly degrade some of the cosmological parameter constraints. In the case that the primordial fluctuations do have a scale-dependent trispectrum, but one follows the standard cosmological analysis without the non-Gaussian covariance term, the inferred cosmological parameters can be significantly biased. 

In the form written in Eq.(\ref{eq:trispectrum}), the isotropic power modulation (non-Gaussian covariance term of Eq.(\ref{eq:Clcov})) is infrared-divergent because of the integral $F_{L=0}(r)\propto \int (dK/K) j_0^2(Kr)$, in which arbitrarily large wavelength modes $(K\rightarrow 0)$ contribute. To study any possible modulation of the isotropic power spectrum, therefore, we need an infrared cutoff. {We also need to assume a form for the power spectrum on large scales; we simply take the near scale-invariant form of primordial power spectrum to be valid at all scales above the infrared cutoff scale.} For $K_{\rm min} r \ll 1$, and $n_s\approx 1$,  $F_0(r)$ can be approximated as
\begin{align}
   F_0(r) &\approx  4 \pi A_\Phi \int_{K_{\rm min}}^\infty \frac{dK}{K} j_0(K r)^2 \approx 4 \pi A_\Phi \ln \left(\frac{1}{K_{\rm min} r}\right)
\end{align}

For $\ell, \ell' \gg 2$, the non-Gaussian power spectrum covariance term is given by
\begin{align}
\cov_{\rm NG}(\hat{C}_\ell, \hat{C}_{\ell'}) \approx & \frac{\taunl A_\Phi \Nextra}{\pi} \int dr_1 r_1^2 \alpha_\ell(r_1, n) \beta_\ell(r_1) \nonumber \\ & \int dr_2 r_2^2 \alpha_{\ell'}(r_2, n) \beta_{\ell'}(r_2),
\label{eq:ngClcov}
\end{align}
where we have defined 
\[\Nextra = \ln\left(\frac{1}{K_{\rm min}r_{\rm cmb}}\right)\] that determines the 
strength of the non-Gaussian covariance. The scale dependence can be 
independently constrained by higher-order modulations for a non-zero $\taunl$. 
In Figure \ref{fig:ngclcov}, we plot the fractional non-Gaussian contribution 
for our fiducial model, for a chosen value of $\Nextra$.

Scale-dependent non-Gaussianity can significantly change the scale dependence of the observed power spectrum and introduce additional cosmic variance uncertainty in the observed spectral index $n_s$ \cite{Bramante:2013moa}. A CMB data analysis allowing for non-Gaussian covariance structure as in Eq.(\ref{eq:Clcov}) will account for the additional uncertainty. Ignoring the non-Gaussian term will result in inference of cosmological parameters that are different than the true values. We provide an example next.

An increasing correlation between $C_\ell$s at large scales can explain \cite{Adhikari2016} the observed power deficit at low multipoles \cite{Ade:2013kta} in the temperature fluctuations. If the low-$\ell$ deficit is due a scale-dependent trispectrum similar to our fiducial model, the decreasing strength of correlations at larger multipoles --- as can be seen in Figure \ref{fig:ngclcov} --- means that the inferred spectral index $n_s$ is shifted higher than the true value. The allowed level of shift due to non-Gaussianity can be much larger at $\Delta n_s \sim 0.04$ \cite{Bramante:2013moa} for example, than the precision of the measurement from the {\it Planck} mission. Specifically, a typical realization of our fiducial non-Gaussian model with the additional parameter $\Nextra=40$ fixed can have large-scale power lower than the true value; in that case, a CMB analysis without the non-Gaussian covariance will produce a biased high estimate of the spectral index as exemplified on the right panel of Figure \ref{fig:ngclcov}. 

In such a scenario, the {\it Planck}-inferred values for other cosmological parameters may also be biased. {Perhaps most straightforwardly, an incorrect reconstruction of the scalar power amplitude will change the inferred bound on (or, in the event of a detection, the value of) the tensor-to-scalar ratio, $r$. This can affect conclusions drawn about the inflaton field range.} However, more generally the structure of non-Gaussian covariance is complex and it is difficult to predict if and how other non-primordial parameters are biased as there are degeneracies between the parameters. A detail analysis of this possibility and how well the parameters $\mathcal{N}, \taunl, n$ in Eq.(\ref{eq:Clcov}) can be constrained by combining power spectrum and higher-order modulations using {\it Planck}, CMB measurements at smaller scales \cite{Aylor:2017haa}, and other large-scale structure probes is left for future study.

\section{Summary and Conclusion} \label{sec:summary}
We have systematically studied the modulation effects of scale-dependent primordial non-Gaussianity in the cosmic microwave background fluctuations. We do so, in detail, by using a scale-dependent local-type trispectrum which has a large collapsed-limit signal i.e. in which long-wavelength modes are significantly coupled to small-scale modes. We assume global statistical isotropy and compute covariances of statistical anisotropy estimators of two-point functions, in the presence of a primordial trispectrum. Such a method is necessary when we want to include multiple observables (temperature and polarization fluctuations for example), that probe the same underlying primordial density field which are additionally correlated in the non-Gaussian model.

While {current constraints on a scale-dependent trispectrum are rather weak,}  we find that for our fiducial model parameters which can explain the hemispherical power asymmetry, the prospects of detection using {\it Planck} data (dipole modulation estimators only) are promising: an estimated log-likelihood improvement of $\sim 11$ using {\it Planck} T+E data up to $\ell_{\rm max}=1000$, with only two extra parameters. Addition of higher-order modulations $L=2,3,4$ improves the signal by $\sim 30\%$. 

A primordial trispectrum generically also produces covariance between different multipoles of the angular power spectrum. If such a scale-dependent non-Gaussian covariance term is present but ignored in CMB analysis, we have shown that the resulting level of bias in the spectral index can be significant. Further, the bias in the spectral index itself is scale dependent, which presents the possibility of its detection by combining small-scale CMB measurements to that of the larger-scale {\it Planck} data. Given that we get much of our precision cosmology from the CMB $C_\ell$s, we must, therefore, constrain primordial trispectra (with possible scale dependence) that have large signal in the collapsed limit.

In addition to constraining the scale-dependent trispectrum parameters from CMB data, there are several other interesting and useful future studies that can extend our work. First, it will be interesting to study how a non-Gaussian covariance from a scale-dependent trispectrum will bias other cosmological parameters and if it can explain some of the current parameter tensions observed from large-scale and small-scale measurements. Second, for ease of numerical evaluations, we mostly focused on multipoles $\ell \geq 30$; a natural extension of this work, therefore, will be to carefully examine the effects of a scale-dependent trispectrum at lower multipoles as the contributing trispectrum configurations begin to deviate from the exact collapsed limit of the trispectrum. Third, it will be useful to compute the consequences of a scale-dependent trispectrum and forecast constraints by including other probes such as CMB lensing and large-scale structure. 

\begin{acknowledgments}
    S.A. is supported by NASA under Contract No. 14-ATP14-0005. S.S. and A.-S.D are supported by the National Science Foundation under Award No. PHY-1719991. This work used the Extreme Science and Engineering Discovery Environment (XSEDE), which is supported by National Science Foundation Grant No. ACI-1548562. We thank Dragan Huterer for discussions and suggesting improvements on the draft.
\end{acknowledgments}

\bibliography{paper_draft.bib}

\begin{thebibliography}{57}%
\makeatletter
\providecommand \@ifxundefined [1]{%
 \@ifx{#1\undefined}
}%
\providecommand \@ifnum [1]{%
 \ifnum #1\expandafter \@firstoftwo
 \else \expandafter \@secondoftwo
 \fi
}%
\providecommand \@ifx [1]{%
 \ifx #1\expandafter \@firstoftwo
 \else \expandafter \@secondoftwo
 \fi
}%
\providecommand \natexlab [1]{#1}%
\providecommand \enquote  [1]{``#1''}%
\providecommand \bibnamefont  [1]{#1}%
\providecommand \bibfnamefont [1]{#1}%
\providecommand \citenamefont [1]{#1}%
\providecommand \href@noop [0]{\@secondoftwo}%
\providecommand \href [0]{\begingroup \@sanitize@url \@href}%
\providecommand \@href[1]{\@@startlink{#1}\@@href}%
\providecommand \@@href[1]{\endgroup#1\@@endlink}%
\providecommand \@sanitize@url [0]{\catcode `\\12\catcode `\$12\catcode
  `\&12\catcode `\#12\catcode `\^12\catcode `\_12\catcode `\%12\relax}%
\providecommand \@@startlink[1]{}%
\providecommand \@@endlink[0]{}%
\providecommand \url  [0]{\begingroup\@sanitize@url \@url }%
\providecommand \@url [1]{\endgroup\@href {#1}{\urlprefix }}%
\providecommand \urlprefix  [0]{URL }%
\providecommand \Eprint [0]{\href }%
\providecommand \doibase [0]{http://dx.doi.org/}%
\providecommand \selectlanguage [0]{\@gobble}%
\providecommand \bibinfo  [0]{\@secondoftwo}%
\providecommand \bibfield  [0]{\@secondoftwo}%
\providecommand \translation [1]{[#1]}%
\providecommand \BibitemOpen [0]{}%
\providecommand \bibitemStop [0]{}%
\providecommand \bibitemNoStop [0]{.\EOS\space}%
\providecommand \EOS [0]{\spacefactor3000\relax}%
\providecommand \BibitemShut  [1]{\csname bibitem#1\endcsname}%
\let\auto@bib@innerbib\@empty
\bibitem [{\citenamefont {Schwarz}\ \emph {et~al.}(2016)\citenamefont
  {Schwarz}, \citenamefont {Copi}, \citenamefont {Huterer},\ and\ \citenamefont
  {Starkman}}]{Schwarz:2015cma}%
  \BibitemOpen
  \bibfield  {author} {\bibinfo {author} {\bibfnamefont {D.~J.}\ \bibnamefont
  {Schwarz}}, \bibinfo {author} {\bibfnamefont {C.~J.}\ \bibnamefont {Copi}},
  \bibinfo {author} {\bibfnamefont {D.}~\bibnamefont {Huterer}}, \ and\
  \bibinfo {author} {\bibfnamefont {G.~D.}\ \bibnamefont {Starkman}},\ }\href
  {\doibase 10.1088/0264-9381/33/18/184001} {\bibfield  {journal} {\bibinfo
  {journal} {Class. Quant. Grav.}\ }\textbf {\bibinfo {volume} {33}},\ \bibinfo
  {pages} {184001} (\bibinfo {year} {2016})},\ \Eprint
  {http://arxiv.org/abs/1510.07929} {arXiv:1510.07929 [astro-ph.CO]}
  \BibitemShut {NoStop}%
\bibitem [{\citenamefont {Erickcek}\ \emph {et~al.}(2008)\citenamefont
  {Erickcek}, \citenamefont {Kamionkowski},\ and\ \citenamefont
  {Carroll}}]{Erickcek:2008sm}%
  \BibitemOpen
  \bibfield  {author} {\bibinfo {author} {\bibfnamefont {A.~L.}\ \bibnamefont
  {Erickcek}}, \bibinfo {author} {\bibfnamefont {M.}~\bibnamefont
  {Kamionkowski}}, \ and\ \bibinfo {author} {\bibfnamefont {S.~M.}\
  \bibnamefont {Carroll}},\ }\href {\doibase 10.1103/PhysRevD.78.123520}
  {\bibfield  {journal} {\bibinfo  {journal} {Phys. Rev.}\ }\textbf {\bibinfo
  {volume} {D78}},\ \bibinfo {pages} {123520} (\bibinfo {year} {2008})},\
  \Eprint {http://arxiv.org/abs/0806.0377} {arXiv:0806.0377 [astro-ph]}
  \BibitemShut {NoStop}%
\bibitem [{\citenamefont {Erickcek}\ \emph {et~al.}(2009)\citenamefont
  {Erickcek}, \citenamefont {Hirata},\ and\ \citenamefont
  {Kamionkowski}}]{Erickcek:2009at}%
  \BibitemOpen
  \bibfield  {author} {\bibinfo {author} {\bibfnamefont {A.~L.}\ \bibnamefont
  {Erickcek}}, \bibinfo {author} {\bibfnamefont {C.~M.}\ \bibnamefont
  {Hirata}}, \ and\ \bibinfo {author} {\bibfnamefont {M.}~\bibnamefont
  {Kamionkowski}},\ }\href {\doibase 10.1103/PhysRevD.80.083507} {\bibfield
  {journal} {\bibinfo  {journal} {Phys. Rev.}\ }\textbf {\bibinfo {volume}
  {D80}},\ \bibinfo {pages} {083507} (\bibinfo {year} {2009})},\ \Eprint
  {http://arxiv.org/abs/0907.0705} {arXiv:0907.0705 [astro-ph.CO]} \BibitemShut
  {NoStop}%
\bibitem [{\citenamefont {Lyth}(2013)}]{Lyth:2013vha}%
  \BibitemOpen
  \bibfield  {author} {\bibinfo {author} {\bibfnamefont {D.~H.}\ \bibnamefont
  {Lyth}},\ }\href {\doibase 10.1088/1475-7516/2013/08/007} {\bibfield
  {journal} {\bibinfo  {journal} {JCAP}\ }\textbf {\bibinfo {volume} {1308}},\
  \bibinfo {pages} {007} (\bibinfo {year} {2013})},\ \Eprint
  {http://arxiv.org/abs/1304.1270} {arXiv:1304.1270 [astro-ph.CO]} \BibitemShut
  {NoStop}%
\bibitem [{\citenamefont {Liddle}\ and\ \citenamefont
  {Cortês}(2013)}]{Liddle:2013czu}%
  \BibitemOpen
  \bibfield  {author} {\bibinfo {author} {\bibfnamefont {A.~R.}\ \bibnamefont
  {Liddle}}\ and\ \bibinfo {author} {\bibfnamefont {M.}~\bibnamefont
  {Cortês}},\ }\href {\doibase 10.1103/PhysRevLett.111.111302} {\bibfield
  {journal} {\bibinfo  {journal} {Phys. Rev. Lett.}\ }\textbf {\bibinfo
  {volume} {111}},\ \bibinfo {pages} {111302} (\bibinfo {year} {2013})},\
  \Eprint {http://arxiv.org/abs/1306.5698} {arXiv:1306.5698 [astro-ph.CO]}
  \BibitemShut {NoStop}%
\bibitem [{\citenamefont {Byrnes}\ \emph
  {et~al.}(2016{\natexlab{a}})\citenamefont {Byrnes}, \citenamefont
  {Domènech}, \citenamefont {Sasaki},\ and\ \citenamefont
  {Takahashi}}]{Byrnes:2016uqw}%
  \BibitemOpen
  \bibfield  {author} {\bibinfo {author} {\bibfnamefont {C.}~\bibnamefont
  {Byrnes}}, \bibinfo {author} {\bibfnamefont {G.}~\bibnamefont {Domènech}},
  \bibinfo {author} {\bibfnamefont {M.}~\bibnamefont {Sasaki}}, \ and\ \bibinfo
  {author} {\bibfnamefont {T.}~\bibnamefont {Takahashi}},\ }\href {\doibase
  10.1088/1475-7516/2016/12/020} {\bibfield  {journal} {\bibinfo  {journal}
  {JCAP}\ }\textbf {\bibinfo {volume} {1612}},\ \bibinfo {pages} {020}
  (\bibinfo {year} {2016}{\natexlab{a}})},\ \Eprint
  {http://arxiv.org/abs/1610.02650} {arXiv:1610.02650 [astro-ph.CO]}
  \BibitemShut {NoStop}%
\bibitem [{\citenamefont {Dai}\ \emph {et~al.}(2013{\natexlab{a}})\citenamefont
  {Dai}, \citenamefont {Jeong}, \citenamefont {Kamionkowski},\ and\
  \citenamefont {Chluba}}]{Dai:2013kfa}%
  \BibitemOpen
  \bibfield  {author} {\bibinfo {author} {\bibfnamefont {L.}~\bibnamefont
  {Dai}}, \bibinfo {author} {\bibfnamefont {D.}~\bibnamefont {Jeong}}, \bibinfo
  {author} {\bibfnamefont {M.}~\bibnamefont {Kamionkowski}}, \ and\ \bibinfo
  {author} {\bibfnamefont {J.}~\bibnamefont {Chluba}},\ }\href {\doibase
  10.1103/PhysRevD.87.123005} {\bibfield  {journal} {\bibinfo  {journal} {Phys.
  Rev.}\ }\textbf {\bibinfo {volume} {D87}},\ \bibinfo {pages} {123005}
  (\bibinfo {year} {2013}{\natexlab{a}})},\ \Eprint
  {http://arxiv.org/abs/1303.6949} {arXiv:1303.6949 [astro-ph.CO]} \BibitemShut
  {NoStop}%
\bibitem [{\citenamefont {Ringeval}\ \emph {et~al.}(2016)\citenamefont
  {Ringeval}, \citenamefont {Yamauchi}, \citenamefont {Yokoyama},\ and\
  \citenamefont {Bouchet}}]{Ringeval:2015ywa}%
  \BibitemOpen
  \bibfield  {author} {\bibinfo {author} {\bibfnamefont {C.}~\bibnamefont
  {Ringeval}}, \bibinfo {author} {\bibfnamefont {D.}~\bibnamefont {Yamauchi}},
  \bibinfo {author} {\bibfnamefont {J.}~\bibnamefont {Yokoyama}}, \ and\
  \bibinfo {author} {\bibfnamefont {F.~R.}\ \bibnamefont {Bouchet}},\ }\href
  {\doibase 10.1088/1475-7516/2016/02/033} {\bibfield  {journal} {\bibinfo
  {journal} {JCAP}\ }\textbf {\bibinfo {volume} {1602}},\ \bibinfo {pages}
  {033} (\bibinfo {year} {2016})},\ \Eprint {http://arxiv.org/abs/1510.01916}
  {arXiv:1510.01916 [astro-ph.CO]} \BibitemShut {NoStop}%
\bibitem [{\citenamefont {Kanno}\ \emph {et~al.}(2013)\citenamefont {Kanno},
  \citenamefont {Sasaki},\ and\ \citenamefont {Tanaka}}]{Kanno:2013ohv}%
  \BibitemOpen
  \bibfield  {author} {\bibinfo {author} {\bibfnamefont {S.}~\bibnamefont
  {Kanno}}, \bibinfo {author} {\bibfnamefont {M.}~\bibnamefont {Sasaki}}, \
  and\ \bibinfo {author} {\bibfnamefont {T.}~\bibnamefont {Tanaka}},\ }\href
  {\doibase 10.1093/ptep/ptt093} {\bibfield  {journal} {\bibinfo  {journal}
  {PTEP}\ }\textbf {\bibinfo {volume} {2013}},\ \bibinfo {pages} {111E01}
  (\bibinfo {year} {2013})},\ \Eprint {http://arxiv.org/abs/1309.1350}
  {arXiv:1309.1350 [astro-ph.CO]} \BibitemShut {NoStop}%
\bibitem [{\citenamefont {Namjoo}\ \emph {et~al.}(2014)\citenamefont {Namjoo},
  \citenamefont {Abolhasani}, \citenamefont {Baghram},\ and\ \citenamefont
  {Firouzjahi}}]{Namjoo:2014nra}%
  \BibitemOpen
  \bibfield  {author} {\bibinfo {author} {\bibfnamefont {M.~H.}\ \bibnamefont
  {Namjoo}}, \bibinfo {author} {\bibfnamefont {A.~A.}\ \bibnamefont
  {Abolhasani}}, \bibinfo {author} {\bibfnamefont {S.}~\bibnamefont {Baghram}},
  \ and\ \bibinfo {author} {\bibfnamefont {H.}~\bibnamefont {Firouzjahi}},\
  }\href {\doibase 10.1088/1475-7516/2014/08/002} {\bibfield  {journal}
  {\bibinfo  {journal} {JCAP}\ }\textbf {\bibinfo {volume} {1408}},\ \bibinfo
  {pages} {002} (\bibinfo {year} {2014})},\ \Eprint
  {http://arxiv.org/abs/1405.7317} {arXiv:1405.7317 [astro-ph.CO]} \BibitemShut
  {NoStop}%
\bibitem [{\citenamefont {Kenton}\ \emph {et~al.}(2015)\citenamefont {Kenton},
  \citenamefont {Mulryne},\ and\ \citenamefont {Thomas}}]{Kenton:2015jga}%
  \BibitemOpen
  \bibfield  {author} {\bibinfo {author} {\bibfnamefont {Z.}~\bibnamefont
  {Kenton}}, \bibinfo {author} {\bibfnamefont {D.~J.}\ \bibnamefont {Mulryne}},
  \ and\ \bibinfo {author} {\bibfnamefont {S.}~\bibnamefont {Thomas}},\ }\href
  {\doibase 10.1103/PhysRevD.92.023505} {\bibfield  {journal} {\bibinfo
  {journal} {Phys. Rev.}\ }\textbf {\bibinfo {volume} {D92}},\ \bibinfo {pages}
  {023505} (\bibinfo {year} {2015})},\ \Eprint
  {http://arxiv.org/abs/1504.05736} {arXiv:1504.05736 [astro-ph.CO]}
  \BibitemShut {NoStop}%
\bibitem [{\citenamefont {Byrnes}\ and\ \citenamefont
  {Tarrant}(2015)}]{Byrnes:2015asa}%
  \BibitemOpen
  \bibfield  {author} {\bibinfo {author} {\bibfnamefont {C.~T.}\ \bibnamefont
  {Byrnes}}\ and\ \bibinfo {author} {\bibfnamefont {E.~R.~M.}\ \bibnamefont
  {Tarrant}},\ }\href {\doibase 10.1088/1475-7516/2015/07/007} {\bibfield
  {journal} {\bibinfo  {journal} {JCAP}\ }\textbf {\bibinfo {volume} {1507}},\
  \bibinfo {pages} {007} (\bibinfo {year} {2015})},\ \Eprint
  {http://arxiv.org/abs/1502.07339} {arXiv:1502.07339 [astro-ph.CO]}
  \BibitemShut {NoStop}%
\bibitem [{\citenamefont {Ade}\ \emph {et~al.}(2016{\natexlab{a}})\citenamefont
  {Ade} \emph {et~al.}}]{PlanckIsotropy2015}%
  \BibitemOpen
  \bibfield  {author} {\bibinfo {author} {\bibfnamefont {P.~A.~R.}\
  \bibnamefont {Ade}} \emph {et~al.} (\bibinfo {collaboration} {Planck}),\
  }\href {\doibase 10.1051/0004-6361/201526681} {\bibfield  {journal} {\bibinfo
   {journal} {Astron. Astrophys.}\ }\textbf {\bibinfo {volume} {594}},\
  \bibinfo {pages} {A16} (\bibinfo {year} {2016}{\natexlab{a}})},\ \Eprint
  {http://arxiv.org/abs/1506.07135} {arXiv:1506.07135 [astro-ph.CO]}
  \BibitemShut {NoStop}%
\bibitem [{\citenamefont {Eriksen}\ \emph {et~al.}(2007)\citenamefont
  {Eriksen}, \citenamefont {Banday}, \citenamefont {Gorski}, \citenamefont
  {Hansen},\ and\ \citenamefont {Lilje}}]{Eriksen:2007pc}%
  \BibitemOpen
  \bibfield  {author} {\bibinfo {author} {\bibfnamefont {H.~K.}\ \bibnamefont
  {Eriksen}}, \bibinfo {author} {\bibfnamefont {A.~J.}\ \bibnamefont {Banday}},
  \bibinfo {author} {\bibfnamefont {K.~M.}\ \bibnamefont {Gorski}}, \bibinfo
  {author} {\bibfnamefont {F.~K.}\ \bibnamefont {Hansen}}, \ and\ \bibinfo
  {author} {\bibfnamefont {P.~B.}\ \bibnamefont {Lilje}},\ }\href {\doibase
  10.1086/518091} {\bibfield  {journal} {\bibinfo  {journal} {Astrophys. J.}\
  }\textbf {\bibinfo {volume} {660}},\ \bibinfo {pages} {L81} (\bibinfo {year}
  {2007})},\ \Eprint {http://arxiv.org/abs/astro-ph/0701089}
  {arXiv:astro-ph/0701089 [astro-ph]} \BibitemShut {NoStop}%
\bibitem [{\citenamefont {Akrami}\ \emph {et~al.}(2014)\citenamefont {Akrami},
  \citenamefont {Fantaye}, \citenamefont {Shafieloo}, \citenamefont {Eriksen},
  \citenamefont {Hansen}, \citenamefont {Banday},\ and\ \citenamefont
  {Górski}}]{Akrami:2014eta}%
  \BibitemOpen
  \bibfield  {author} {\bibinfo {author} {\bibfnamefont {Y.}~\bibnamefont
  {Akrami}}, \bibinfo {author} {\bibfnamefont {Y.}~\bibnamefont {Fantaye}},
  \bibinfo {author} {\bibfnamefont {A.}~\bibnamefont {Shafieloo}}, \bibinfo
  {author} {\bibfnamefont {H.~K.}\ \bibnamefont {Eriksen}}, \bibinfo {author}
  {\bibfnamefont {F.~K.}\ \bibnamefont {Hansen}}, \bibinfo {author}
  {\bibfnamefont {A.~J.}\ \bibnamefont {Banday}}, \ and\ \bibinfo {author}
  {\bibfnamefont {K.~M.}\ \bibnamefont {Górski}},\ }\href {\doibase
  10.1088/2041-8205/784/2/L42} {\bibfield  {journal} {\bibinfo  {journal}
  {Astrophys. J.}\ }\textbf {\bibinfo {volume} {784}},\ \bibinfo {pages} {L42}
  (\bibinfo {year} {2014})},\ \Eprint {http://arxiv.org/abs/1402.0870}
  {arXiv:1402.0870 [astro-ph.CO]} \BibitemShut {NoStop}%
\bibitem [{\citenamefont {Adhikari}(2015)}]{Adhikari:2014mua}%
  \BibitemOpen
  \bibfield  {author} {\bibinfo {author} {\bibfnamefont {S.}~\bibnamefont
  {Adhikari}},\ }\href {\doibase 10.1093/mnras/stu2408} {\bibfield  {journal}
  {\bibinfo  {journal} {Mon. Not. Roy. Astron. Soc.}\ }\textbf {\bibinfo
  {volume} {446}},\ \bibinfo {pages} {4232} (\bibinfo {year} {2015})},\ \Eprint
  {http://arxiv.org/abs/1408.5396} {arXiv:1408.5396 [astro-ph.CO]} \BibitemShut
  {NoStop}%
\bibitem [{\citenamefont {Contreras}\ \emph {et~al.}(2017)\citenamefont
  {Contreras}, \citenamefont {Zibin}, \citenamefont {Scott}, \citenamefont
  {Banday},\ and\ \citenamefont {Górski}}]{Contreras2017}%
  \BibitemOpen
  \bibfield  {author} {\bibinfo {author} {\bibfnamefont {D.}~\bibnamefont
  {Contreras}}, \bibinfo {author} {\bibfnamefont {J.~P.}\ \bibnamefont
  {Zibin}}, \bibinfo {author} {\bibfnamefont {D.}~\bibnamefont {Scott}},
  \bibinfo {author} {\bibfnamefont {A.~J.}\ \bibnamefont {Banday}}, \ and\
  \bibinfo {author} {\bibfnamefont {K.~M.}\ \bibnamefont {Górski}},\ }\href
  {\doibase 10.1103/PhysRevD.96.123522} {\bibfield  {journal} {\bibinfo
  {journal} {Phys. Rev.}\ }\textbf {\bibinfo {volume} {D96}},\ \bibinfo {pages}
  {123522} (\bibinfo {year} {2017})},\ \Eprint
  {http://arxiv.org/abs/1704.03143} {arXiv:1704.03143 [astro-ph.CO]}
  \BibitemShut {NoStop}%
\bibitem [{\citenamefont {Zhai}\ and\ \citenamefont
  {Blanton}(2017)}]{Zhai:2017ibd}%
  \BibitemOpen
  \bibfield  {author} {\bibinfo {author} {\bibfnamefont {Z.}~\bibnamefont
  {Zhai}}\ and\ \bibinfo {author} {\bibfnamefont {M.}~\bibnamefont {Blanton}},\
  }\href {\doibase 10.3847/1538-4357/aa93e1} {\bibfield  {journal} {\bibinfo
  {journal} {Astrophys. J.}\ }\textbf {\bibinfo {volume} {850}},\ \bibinfo
  {pages} {41} (\bibinfo {year} {2017})},\ \Eprint
  {http://arxiv.org/abs/1707.06555} {arXiv:1707.06555 [astro-ph.CO]}
  \BibitemShut {NoStop}%
\bibitem [{\citenamefont {Terrana}\ \emph {et~al.}(2017)\citenamefont
  {Terrana}, \citenamefont {Harris},\ and\ \citenamefont
  {Johnson}}]{Terrana2016}%
  \BibitemOpen
  \bibfield  {author} {\bibinfo {author} {\bibfnamefont {A.}~\bibnamefont
  {Terrana}}, \bibinfo {author} {\bibfnamefont {M.-J.}\ \bibnamefont {Harris}},
  \ and\ \bibinfo {author} {\bibfnamefont {M.~C.}\ \bibnamefont {Johnson}},\
  }\href {\doibase 10.1088/1475-7516/2017/02/040} {\bibfield  {journal}
  {\bibinfo  {journal} {Journal of Cosmology and Astroparticle Physics}\
  }\textbf {\bibinfo {volume} {2017}},\ \bibinfo {pages} {040} (\bibinfo {year}
  {2017})},\ \Eprint {http://arxiv.org/abs/1610.06919} {arXiv:1610.06919}
  \BibitemShut {NoStop}%
\bibitem [{\citenamefont {Deutsch}\ \emph
  {et~al.}(2017{\natexlab{a}})\citenamefont {Deutsch}, \citenamefont {Johnson},
  \citenamefont {M{\"{u}}nchmeyer},\ and\ \citenamefont
  {Terrana}}]{Deutsch2017a}%
  \BibitemOpen
  \bibfield  {author} {\bibinfo {author} {\bibfnamefont {A.-S.}\ \bibnamefont
  {Deutsch}}, \bibinfo {author} {\bibfnamefont {M.~C.}\ \bibnamefont
  {Johnson}}, \bibinfo {author} {\bibfnamefont {M.}~\bibnamefont
  {M{\"{u}}nchmeyer}}, \ and\ \bibinfo {author} {\bibfnamefont
  {A.}~\bibnamefont {Terrana}},\ }\href {http://arxiv.org/abs/1705.08907} {\
  (\bibinfo {year} {2017}{\natexlab{a}})},\ \Eprint
  {http://arxiv.org/abs/1705.08907} {arXiv:1705.08907} \BibitemShut {NoStop}%
\bibitem [{\citenamefont {Deutsch}\ \emph
  {et~al.}(2017{\natexlab{b}})\citenamefont {Deutsch}, \citenamefont
  {Dimastrogiovanni}, \citenamefont {Johnson}, \citenamefont
  {M{\"{u}}nchmeyer},\ and\ \citenamefont {Terrana}}]{Deutsch2017}%
  \BibitemOpen
  \bibfield  {author} {\bibinfo {author} {\bibfnamefont {A.-S.}\ \bibnamefont
  {Deutsch}}, \bibinfo {author} {\bibfnamefont {E.}~\bibnamefont
  {Dimastrogiovanni}}, \bibinfo {author} {\bibfnamefont {M.~C.}\ \bibnamefont
  {Johnson}}, \bibinfo {author} {\bibfnamefont {M.}~\bibnamefont
  {M{\"{u}}nchmeyer}}, \ and\ \bibinfo {author} {\bibfnamefont
  {A.}~\bibnamefont {Terrana}},\ }\href {http://arxiv.org/abs/1707.08129} {\
  (\bibinfo {year} {2017}{\natexlab{b}})},\ \Eprint
  {http://arxiv.org/abs/1707.08129} {arXiv:1707.08129} \BibitemShut {NoStop}%
\bibitem [{\citenamefont {Shiraishi}\ \emph {et~al.}(2016)\citenamefont
  {Shiraishi}, \citenamefont {Muñoz}, \citenamefont {Kamionkowski},\ and\
  \citenamefont {Raccanelli}}]{Shiraishi:2016omb}%
  \BibitemOpen
  \bibfield  {author} {\bibinfo {author} {\bibfnamefont {M.}~\bibnamefont
  {Shiraishi}}, \bibinfo {author} {\bibfnamefont {J.~B.}\ \bibnamefont
  {Muñoz}}, \bibinfo {author} {\bibfnamefont {M.}~\bibnamefont
  {Kamionkowski}}, \ and\ \bibinfo {author} {\bibfnamefont {A.}~\bibnamefont
  {Raccanelli}},\ }\href {\doibase 10.1103/PhysRevD.93.103506} {\bibfield
  {journal} {\bibinfo  {journal} {Phys. Rev.}\ }\textbf {\bibinfo {volume}
  {D93}},\ \bibinfo {pages} {103506} (\bibinfo {year} {2016})},\ \Eprint
  {http://arxiv.org/abs/1603.01206} {arXiv:1603.01206 [astro-ph.CO]}
  \BibitemShut {NoStop}%
\bibitem [{\citenamefont {Jazayeri}\ \emph {et~al.}(2014)\citenamefont
  {Jazayeri}, \citenamefont {Akrami}, \citenamefont {Firouzjahi}, \citenamefont
  {Solomon},\ and\ \citenamefont {Wang}}]{Jazayeri:2014nya}%
  \BibitemOpen
  \bibfield  {author} {\bibinfo {author} {\bibfnamefont {S.}~\bibnamefont
  {Jazayeri}}, \bibinfo {author} {\bibfnamefont {Y.}~\bibnamefont {Akrami}},
  \bibinfo {author} {\bibfnamefont {H.}~\bibnamefont {Firouzjahi}}, \bibinfo
  {author} {\bibfnamefont {A.~R.}\ \bibnamefont {Solomon}}, \ and\ \bibinfo
  {author} {\bibfnamefont {Y.}~\bibnamefont {Wang}},\ }\href {\doibase
  10.1088/1475-7516/2014/11/044} {\bibfield  {journal} {\bibinfo  {journal}
  {JCAP}\ }\textbf {\bibinfo {volume} {1411}},\ \bibinfo {pages} {044}
  (\bibinfo {year} {2014})},\ \Eprint {http://arxiv.org/abs/1408.3057}
  {arXiv:1408.3057 [astro-ph.CO]} \BibitemShut {NoStop}%
\bibitem [{\citenamefont {Gordon}\ \emph {et~al.}(2005)\citenamefont {Gordon},
  \citenamefont {Hu}, \citenamefont {Huterer},\ and\ \citenamefont
  {Crawford}}]{Gordon:2005ai}%
  \BibitemOpen
  \bibfield  {author} {\bibinfo {author} {\bibfnamefont {C.}~\bibnamefont
  {Gordon}}, \bibinfo {author} {\bibfnamefont {W.}~\bibnamefont {Hu}}, \bibinfo
  {author} {\bibfnamefont {D.}~\bibnamefont {Huterer}}, \ and\ \bibinfo
  {author} {\bibfnamefont {T.~M.}\ \bibnamefont {Crawford}},\ }\href {\doibase
  10.1103/PhysRevD.72.103002} {\bibfield  {journal} {\bibinfo  {journal} {Phys.
  Rev.}\ }\textbf {\bibinfo {volume} {D72}},\ \bibinfo {pages} {103002}
  (\bibinfo {year} {2005})},\ \Eprint {http://arxiv.org/abs/astro-ph/0509301}
  {arXiv:astro-ph/0509301 [astro-ph]} \BibitemShut {NoStop}%
\bibitem [{\citenamefont {Adhikari}\ \emph {et~al.}(2016)\citenamefont
  {Adhikari}, \citenamefont {Shandera},\ and\ \citenamefont
  {Erickcek}}]{Adhikari2016}%
  \BibitemOpen
  \bibfield  {author} {\bibinfo {author} {\bibfnamefont {S.}~\bibnamefont
  {Adhikari}}, \bibinfo {author} {\bibfnamefont {S.}~\bibnamefont {Shandera}},
  \ and\ \bibinfo {author} {\bibfnamefont {A.~L.}\ \bibnamefont {Erickcek}},\
  }\href {\doibase 10.1103/PhysRevD.93.023524} {\bibfield  {journal} {\bibinfo
  {journal} {Physical Review D}\ }\textbf {\bibinfo {volume} {93}},\ \bibinfo
  {pages} {023524} (\bibinfo {year} {2016})},\ \Eprint
  {http://arxiv.org/abs/1508.06489} {arXiv:1508.06489} \BibitemShut {NoStop}%
\bibitem [{\citenamefont {Ferreira}\ and\ \citenamefont
  {Magueijo}(1997)}]{Ferreira:1997wd}%
  \BibitemOpen
  \bibfield  {author} {\bibinfo {author} {\bibfnamefont {P.~G.}\ \bibnamefont
  {Ferreira}}\ and\ \bibinfo {author} {\bibfnamefont {J.}~\bibnamefont
  {Magueijo}},\ }\href {\doibase 10.1103/PhysRevD.56.4578} {\bibfield
  {journal} {\bibinfo  {journal} {Phys. Rev.}\ }\textbf {\bibinfo {volume}
  {D56}},\ \bibinfo {pages} {4578} (\bibinfo {year} {1997})},\ \Eprint
  {http://arxiv.org/abs/astro-ph/9704052} {arXiv:astro-ph/9704052 [astro-ph]}
  \BibitemShut {NoStop}%
\bibitem [{\citenamefont {Hanson}\ and\ \citenamefont
  {Lewis}(2009)}]{Hanson2009}%
  \BibitemOpen
  \bibfield  {author} {\bibinfo {author} {\bibfnamefont {D.}~\bibnamefont
  {Hanson}}\ and\ \bibinfo {author} {\bibfnamefont {A.}~\bibnamefont {Lewis}},\
  }\href {\doibase 10.1103/PhysRevD.80.063004} {\bibfield  {journal} {\bibinfo
  {journal} {Phys. Rev.}\ }\textbf {\bibinfo {volume} {D80}},\ \bibinfo {pages}
  {063004} (\bibinfo {year} {2009})},\ \Eprint {http://arxiv.org/abs/0908.0963}
  {arXiv:0908.0963 [astro-ph.CO]} \BibitemShut {NoStop}%
\bibitem [{\citenamefont {Dai}\ \emph {et~al.}(2013{\natexlab{b}})\citenamefont
  {Dai}, \citenamefont {Jeong},\ and\ \citenamefont
  {Kamionkowski}}]{Dai:2013ikl}%
  \BibitemOpen
  \bibfield  {author} {\bibinfo {author} {\bibfnamefont {L.}~\bibnamefont
  {Dai}}, \bibinfo {author} {\bibfnamefont {D.}~\bibnamefont {Jeong}}, \ and\
  \bibinfo {author} {\bibfnamefont {M.}~\bibnamefont {Kamionkowski}},\ }\href
  {\doibase 10.1103/PhysRevD.87.103006} {\bibfield  {journal} {\bibinfo
  {journal} {Phys. Rev.}\ }\textbf {\bibinfo {volume} {D87}},\ \bibinfo {pages}
  {103006} (\bibinfo {year} {2013}{\natexlab{b}})},\ \Eprint
  {http://arxiv.org/abs/1302.1868} {arXiv:1302.1868 [astro-ph.CO]} \BibitemShut
  {NoStop}%
\bibitem [{\citenamefont {Ade}\ \emph {et~al.}(2014{\natexlab{a}})\citenamefont
  {Ade} \emph {et~al.}}]{Ade:2013ydc}%
  \BibitemOpen
  \bibfield  {author} {\bibinfo {author} {\bibfnamefont {P.~A.~R.}\
  \bibnamefont {Ade}} \emph {et~al.} (\bibinfo {collaboration} {Planck}),\
  }\href {\doibase 10.1051/0004-6361/201321554} {\bibfield  {journal} {\bibinfo
   {journal} {Astron. Astrophys.}\ }\textbf {\bibinfo {volume} {571}},\
  \bibinfo {pages} {A24} (\bibinfo {year} {2014}{\natexlab{a}})},\ \Eprint
  {http://arxiv.org/abs/1303.5084} {arXiv:1303.5084 [astro-ph.CO]} \BibitemShut
  {NoStop}%
\bibitem [{\citenamefont {Feng}\ \emph {et~al.}(2015)\citenamefont {Feng},
  \citenamefont {Cooray}, \citenamefont {Smidt}, \citenamefont {O'Bryan},
  \citenamefont {Keating},\ and\ \citenamefont {Regan}}]{Feng:2015pva}%
  \BibitemOpen
  \bibfield  {author} {\bibinfo {author} {\bibfnamefont {C.}~\bibnamefont
  {Feng}}, \bibinfo {author} {\bibfnamefont {A.}~\bibnamefont {Cooray}},
  \bibinfo {author} {\bibfnamefont {J.}~\bibnamefont {Smidt}}, \bibinfo
  {author} {\bibfnamefont {J.}~\bibnamefont {O'Bryan}}, \bibinfo {author}
  {\bibfnamefont {B.}~\bibnamefont {Keating}}, \ and\ \bibinfo {author}
  {\bibfnamefont {D.}~\bibnamefont {Regan}},\ }\href {\doibase
  10.1103/PhysRevD.92.043509} {\bibfield  {journal} {\bibinfo  {journal} {Phys.
  Rev.}\ }\textbf {\bibinfo {volume} {D92}},\ \bibinfo {pages} {043509}
  (\bibinfo {year} {2015})},\ \Eprint {http://arxiv.org/abs/1502.00585}
  {arXiv:1502.00585 [astro-ph.CO]} \BibitemShut {NoStop}%
\bibitem [{\citenamefont {Regan}\ \emph {et~al.}(2015)\citenamefont {Regan},
  \citenamefont {Gosenca},\ and\ \citenamefont {Seery}}]{Regan:2013jua}%
  \BibitemOpen
  \bibfield  {author} {\bibinfo {author} {\bibfnamefont {D.}~\bibnamefont
  {Regan}}, \bibinfo {author} {\bibfnamefont {M.}~\bibnamefont {Gosenca}}, \
  and\ \bibinfo {author} {\bibfnamefont {D.}~\bibnamefont {Seery}},\ }\href
  {\doibase 10.1088/1475-7516/2015/01/013} {\bibfield  {journal} {\bibinfo
  {journal} {JCAP}\ }\textbf {\bibinfo {volume} {1501}},\ \bibinfo {pages}
  {013} (\bibinfo {year} {2015})},\ \Eprint {http://arxiv.org/abs/1310.8617}
  {arXiv:1310.8617 [astro-ph.CO]} \BibitemShut {NoStop}%
\bibitem [{\citenamefont {Dvorkin}\ \emph {et~al.}(2008)\citenamefont
  {Dvorkin}, \citenamefont {Peiris},\ and\ \citenamefont {Hu}}]{Dvorkin2008}%
  \BibitemOpen
  \bibfield  {author} {\bibinfo {author} {\bibfnamefont {C.}~\bibnamefont
  {Dvorkin}}, \bibinfo {author} {\bibfnamefont {H.~V.}\ \bibnamefont {Peiris}},
  \ and\ \bibinfo {author} {\bibfnamefont {W.}~\bibnamefont {Hu}},\ }\href
  {\doibase 10.1103/PhysRevD.77.063008} {\bibfield  {journal} {\bibinfo
  {journal} {Physical Review D - Particles, Fields, Gravitation and Cosmology}\
  }\textbf {\bibinfo {volume} {77}},\ \bibinfo {pages} {1} (\bibinfo {year}
  {2008})},\ \Eprint {http://arxiv.org/abs/arXiv:0711.2321v2}
  {arXiv:arXiv:0711.2321v2} \BibitemShut {NoStop}%
\bibitem [{\citenamefont {Nelson}\ and\ \citenamefont
  {Shandera}(2013)}]{Nelson:2012sb}%
  \BibitemOpen
  \bibfield  {author} {\bibinfo {author} {\bibfnamefont {E.}~\bibnamefont
  {Nelson}}\ and\ \bibinfo {author} {\bibfnamefont {S.}~\bibnamefont
  {Shandera}},\ }\href {\doibase 10.1103/PhysRevLett.110.131301} {\bibfield
  {journal} {\bibinfo  {journal} {Phys. Rev. Lett.}\ }\textbf {\bibinfo
  {volume} {110}},\ \bibinfo {pages} {131301} (\bibinfo {year} {2013})},\
  \Eprint {http://arxiv.org/abs/1212.4550} {arXiv:1212.4550 [astro-ph.CO]}
  \BibitemShut {NoStop}%
\bibitem [{\citenamefont {LoVerde}\ \emph {et~al.}(2013)\citenamefont
  {LoVerde}, \citenamefont {Nelson},\ and\ \citenamefont
  {Shandera}}]{LoVerde:2013xka}%
  \BibitemOpen
  \bibfield  {author} {\bibinfo {author} {\bibfnamefont {M.}~\bibnamefont
  {LoVerde}}, \bibinfo {author} {\bibfnamefont {E.}~\bibnamefont {Nelson}}, \
  and\ \bibinfo {author} {\bibfnamefont {S.}~\bibnamefont {Shandera}},\ }\href
  {\doibase 10.1088/1475-7516/2013/06/024} {\bibfield  {journal} {\bibinfo
  {journal} {JCAP}\ }\textbf {\bibinfo {volume} {1306}},\ \bibinfo {pages}
  {024} (\bibinfo {year} {2013})},\ \Eprint {http://arxiv.org/abs/1303.3549}
  {arXiv:1303.3549 [astro-ph.CO]} \BibitemShut {NoStop}%
\bibitem [{\citenamefont {Scoccimarro}\ \emph {et~al.}(2012)\citenamefont
  {Scoccimarro}, \citenamefont {Hui}, \citenamefont {Manera},\ and\
  \citenamefont {Chan}}]{Scoccimarro:2011pz}%
  \BibitemOpen
  \bibfield  {author} {\bibinfo {author} {\bibfnamefont {R.}~\bibnamefont
  {Scoccimarro}}, \bibinfo {author} {\bibfnamefont {L.}~\bibnamefont {Hui}},
  \bibinfo {author} {\bibfnamefont {M.}~\bibnamefont {Manera}}, \ and\ \bibinfo
  {author} {\bibfnamefont {K.~C.}\ \bibnamefont {Chan}},\ }\href {\doibase
  10.1103/PhysRevD.85.083002} {\bibfield  {journal} {\bibinfo  {journal} {Phys.
  Rev.}\ }\textbf {\bibinfo {volume} {D85}},\ \bibinfo {pages} {083002}
  (\bibinfo {year} {2012})},\ \Eprint {http://arxiv.org/abs/1108.5512}
  {arXiv:1108.5512 [astro-ph.CO]} \BibitemShut {NoStop}%
\bibitem [{\citenamefont {Baytas}\ \emph {et~al.}(2015)\citenamefont {Baytas},
  \citenamefont {Kesavan}, \citenamefont {Nelson}, \citenamefont {Park},\ and\
  \citenamefont {Shandera}}]{Baytas:2015nja}%
  \BibitemOpen
  \bibfield  {author} {\bibinfo {author} {\bibfnamefont {B.}~\bibnamefont
  {Baytas}}, \bibinfo {author} {\bibfnamefont {A.}~\bibnamefont {Kesavan}},
  \bibinfo {author} {\bibfnamefont {E.}~\bibnamefont {Nelson}}, \bibinfo
  {author} {\bibfnamefont {S.}~\bibnamefont {Park}}, \ and\ \bibinfo {author}
  {\bibfnamefont {S.}~\bibnamefont {Shandera}},\ }\href {\doibase
  10.1103/PhysRevD.91.083518} {\bibfield  {journal} {\bibinfo  {journal} {Phys.
  Rev.}\ }\textbf {\bibinfo {volume} {D91}},\ \bibinfo {pages} {083518}
  (\bibinfo {year} {2015})},\ \Eprint {http://arxiv.org/abs/1502.01009}
  {arXiv:1502.01009 [astro-ph.CO]} \BibitemShut {NoStop}%
\bibitem [{\citenamefont {Suyama}\ and\ \citenamefont
  {Yamaguchi}(2008)}]{Suyama:2007bg}%
  \BibitemOpen
  \bibfield  {author} {\bibinfo {author} {\bibfnamefont {T.}~\bibnamefont
  {Suyama}}\ and\ \bibinfo {author} {\bibfnamefont {M.}~\bibnamefont
  {Yamaguchi}},\ }\href {\doibase 10.1103/PhysRevD.77.023505} {\bibfield
  {journal} {\bibinfo  {journal} {Phys. Rev.}\ }\textbf {\bibinfo {volume}
  {D77}},\ \bibinfo {pages} {023505} (\bibinfo {year} {2008})},\ \Eprint
  {http://arxiv.org/abs/0709.2545} {arXiv:0709.2545 [astro-ph]} \BibitemShut
  {NoStop}%
\bibitem [{\citenamefont {Chen}\ and\ \citenamefont
  {Wang}(2010)}]{Chen:2009zp}%
  \BibitemOpen
  \bibfield  {author} {\bibinfo {author} {\bibfnamefont {X.}~\bibnamefont
  {Chen}}\ and\ \bibinfo {author} {\bibfnamefont {Y.}~\bibnamefont {Wang}},\
  }\href {\doibase 10.1088/1475-7516/2010/04/027} {\bibfield  {journal}
  {\bibinfo  {journal} {JCAP}\ }\textbf {\bibinfo {volume} {1004}},\ \bibinfo
  {pages} {027} (\bibinfo {year} {2010})},\ \Eprint
  {http://arxiv.org/abs/0911.3380} {arXiv:0911.3380 [hep-th]} \BibitemShut
  {NoStop}%
\bibitem [{\citenamefont {Assassi}\ \emph {et~al.}(2012)\citenamefont
  {Assassi}, \citenamefont {Baumann},\ and\ \citenamefont
  {Green}}]{Assassi:2012zq}%
  \BibitemOpen
  \bibfield  {author} {\bibinfo {author} {\bibfnamefont {V.}~\bibnamefont
  {Assassi}}, \bibinfo {author} {\bibfnamefont {D.}~\bibnamefont {Baumann}}, \
  and\ \bibinfo {author} {\bibfnamefont {D.}~\bibnamefont {Green}},\ }\href
  {\doibase 10.1088/1475-7516/2012/11/047} {\bibfield  {journal} {\bibinfo
  {journal} {JCAP}\ }\textbf {\bibinfo {volume} {1211}},\ \bibinfo {pages}
  {047} (\bibinfo {year} {2012})},\ \Eprint {http://arxiv.org/abs/1204.4207}
  {arXiv:1204.4207 [hep-th]} \BibitemShut {NoStop}%
\bibitem [{\citenamefont {Bonga}\ \emph {et~al.}(2016)\citenamefont {Bonga},
  \citenamefont {Brahma}, \citenamefont {Deutsch},\ and\ \citenamefont
  {Shandera}}]{Bonga:2015urq}%
  \BibitemOpen
  \bibfield  {author} {\bibinfo {author} {\bibfnamefont {B.}~\bibnamefont
  {Bonga}}, \bibinfo {author} {\bibfnamefont {S.}~\bibnamefont {Brahma}},
  \bibinfo {author} {\bibfnamefont {A.-S.}\ \bibnamefont {Deutsch}}, \ and\
  \bibinfo {author} {\bibfnamefont {S.}~\bibnamefont {Shandera}},\ }\href
  {\doibase 10.1088/1475-7516/2016/05/018} {\bibfield  {journal} {\bibinfo
  {journal} {JCAP}\ }\textbf {\bibinfo {volume} {1605}},\ \bibinfo {pages}
  {018} (\bibinfo {year} {2016})},\ \Eprint {http://arxiv.org/abs/1512.05365}
  {arXiv:1512.05365 [astro-ph.CO]} \BibitemShut {NoStop}%
\bibitem [{\citenamefont {Deutsch}(2017)}]{Deutsch:2017rsn}%
  \BibitemOpen
  \bibfield  {author} {\bibinfo {author} {\bibfnamefont {A.-S.}\ \bibnamefont
  {Deutsch}},\ }\href@noop {} {\  (\bibinfo {year} {2017})},\ \Eprint
  {http://arxiv.org/abs/1704.01004} {arXiv:1704.01004 [astro-ph.CO]}
  \BibitemShut {NoStop}%
\bibitem [{\citenamefont {Namjoo}\ \emph {et~al.}(2015)\citenamefont {Namjoo},
  \citenamefont {Abolhasani}, \citenamefont {Assadullahi}, \citenamefont
  {Baghram}, \citenamefont {Firouzjahi},\ and\ \citenamefont
  {Wands}}]{Namjoo:2014pqa}%
  \BibitemOpen
  \bibfield  {author} {\bibinfo {author} {\bibfnamefont {M.~H.}\ \bibnamefont
  {Namjoo}}, \bibinfo {author} {\bibfnamefont {A.~A.}\ \bibnamefont
  {Abolhasani}}, \bibinfo {author} {\bibfnamefont {H.}~\bibnamefont
  {Assadullahi}}, \bibinfo {author} {\bibfnamefont {S.}~\bibnamefont
  {Baghram}}, \bibinfo {author} {\bibfnamefont {H.}~\bibnamefont {Firouzjahi}},
  \ and\ \bibinfo {author} {\bibfnamefont {D.}~\bibnamefont {Wands}},\ }\href
  {\doibase 10.1088/1475-7516/2015/05/015} {\bibfield  {journal} {\bibinfo
  {journal} {JCAP}\ }\textbf {\bibinfo {volume} {1505}},\ \bibinfo {pages}
  {015} (\bibinfo {year} {2015})},\ \Eprint {http://arxiv.org/abs/1411.5312}
  {arXiv:1411.5312 [astro-ph.CO]} \BibitemShut {NoStop}%
\bibitem [{\citenamefont {Hu}(2001)}]{Hu:2001fa}%
  \BibitemOpen
  \bibfield  {author} {\bibinfo {author} {\bibfnamefont {W.}~\bibnamefont
  {Hu}},\ }\href {\doibase 10.1103/PhysRevD.64.083005} {\bibfield  {journal}
  {\bibinfo  {journal} {Phys. Rev.}\ }\textbf {\bibinfo {volume} {D64}},\
  \bibinfo {pages} {083005} (\bibinfo {year} {2001})},\ \Eprint
  {http://arxiv.org/abs/astro-ph/0105117} {arXiv:astro-ph/0105117 [astro-ph]}
  \BibitemShut {NoStop}%
\bibitem [{\citenamefont {Okamoto}\ and\ \citenamefont
  {Hu}(2002)}]{Okamoto:2002ik}%
  \BibitemOpen
  \bibfield  {author} {\bibinfo {author} {\bibfnamefont {T.}~\bibnamefont
  {Okamoto}}\ and\ \bibinfo {author} {\bibfnamefont {W.}~\bibnamefont {Hu}},\
  }\href {\doibase 10.1103/PhysRevD.66.063008} {\bibfield  {journal} {\bibinfo
  {journal} {Phys. Rev.}\ }\textbf {\bibinfo {volume} {D66}},\ \bibinfo {pages}
  {063008} (\bibinfo {year} {2002})},\ \Eprint
  {http://arxiv.org/abs/astro-ph/0206155} {arXiv:astro-ph/0206155 [astro-ph]}
  \BibitemShut {NoStop}%
\bibitem [{\citenamefont {Byrnes}\ \emph {et~al.}(2010)\citenamefont {Byrnes},
  \citenamefont {Gerstenlauer}, \citenamefont {Nurmi}, \citenamefont
  {Tasinato},\ and\ \citenamefont {Wands}}]{Byrnes:2010ft}%
  \BibitemOpen
  \bibfield  {author} {\bibinfo {author} {\bibfnamefont {C.~T.}\ \bibnamefont
  {Byrnes}}, \bibinfo {author} {\bibfnamefont {M.}~\bibnamefont
  {Gerstenlauer}}, \bibinfo {author} {\bibfnamefont {S.}~\bibnamefont {Nurmi}},
  \bibinfo {author} {\bibfnamefont {G.}~\bibnamefont {Tasinato}}, \ and\
  \bibinfo {author} {\bibfnamefont {D.}~\bibnamefont {Wands}},\ }\href
  {\doibase 10.1088/1475-7516/2010/10/004} {\bibfield  {journal} {\bibinfo
  {journal} {JCAP}\ }\textbf {\bibinfo {volume} {1010}},\ \bibinfo {pages}
  {004} (\bibinfo {year} {2010})},\ \Eprint {http://arxiv.org/abs/1007.4277}
  {arXiv:1007.4277 [astro-ph.CO]} \BibitemShut {NoStop}%
\bibitem [{\citenamefont {Regan}\ \emph {et~al.}(2010)\citenamefont {Regan},
  \citenamefont {Shellard},\ and\ \citenamefont {Fergusson}}]{Regan:2010cn}%
  \BibitemOpen
  \bibfield  {author} {\bibinfo {author} {\bibfnamefont {D.~M.}\ \bibnamefont
  {Regan}}, \bibinfo {author} {\bibfnamefont {E.~P.~S.}\ \bibnamefont
  {Shellard}}, \ and\ \bibinfo {author} {\bibfnamefont {J.~R.}\ \bibnamefont
  {Fergusson}},\ }\href {\doibase 10.1103/PhysRevD.82.023520} {\bibfield
  {journal} {\bibinfo  {journal} {Phys. Rev.}\ }\textbf {\bibinfo {volume}
  {D82}},\ \bibinfo {pages} {023520} (\bibinfo {year} {2010})},\ \Eprint
  {http://arxiv.org/abs/1004.2915} {arXiv:1004.2915 [astro-ph.CO]} \BibitemShut
  {NoStop}%
\bibitem [{\citenamefont {Galli}\ \emph {et~al.}(2014)\citenamefont {Galli},
  \citenamefont {Benabed}, \citenamefont {Bouchet}, \citenamefont {Cardoso},
  \citenamefont {Elsner}, \citenamefont {Hivon}, \citenamefont {Mangilli},
  \citenamefont {Prunet},\ and\ \citenamefont {Wandelt}}]{Galli:2014kla}%
  \BibitemOpen
  \bibfield  {author} {\bibinfo {author} {\bibfnamefont {S.}~\bibnamefont
  {Galli}}, \bibinfo {author} {\bibfnamefont {K.}~\bibnamefont {Benabed}},
  \bibinfo {author} {\bibfnamefont {F.}~\bibnamefont {Bouchet}}, \bibinfo
  {author} {\bibfnamefont {J.-F.}\ \bibnamefont {Cardoso}}, \bibinfo {author}
  {\bibfnamefont {F.}~\bibnamefont {Elsner}}, \bibinfo {author} {\bibfnamefont
  {E.}~\bibnamefont {Hivon}}, \bibinfo {author} {\bibfnamefont
  {A.}~\bibnamefont {Mangilli}}, \bibinfo {author} {\bibfnamefont
  {S.}~\bibnamefont {Prunet}}, \ and\ \bibinfo {author} {\bibfnamefont
  {B.}~\bibnamefont {Wandelt}},\ }\href {\doibase 10.1103/PhysRevD.90.063504}
  {\bibfield  {journal} {\bibinfo  {journal} {Phys. Rev.}\ }\textbf {\bibinfo
  {volume} {D90}},\ \bibinfo {pages} {063504} (\bibinfo {year} {2014})},\
  \Eprint {http://arxiv.org/abs/1403.5271} {arXiv:1403.5271 [astro-ph.CO]}
  \BibitemShut {NoStop}%
\bibitem [{\citenamefont {Lewis}\ \emph {et~al.}(2000)\citenamefont {Lewis},
  \citenamefont {Challinor},\ and\ \citenamefont {Lasenby}}]{Lewis:1999bs}%
  \BibitemOpen
  \bibfield  {author} {\bibinfo {author} {\bibfnamefont {A.}~\bibnamefont
  {Lewis}}, \bibinfo {author} {\bibfnamefont {A.}~\bibnamefont {Challinor}}, \
  and\ \bibinfo {author} {\bibfnamefont {A.}~\bibnamefont {Lasenby}},\ }\href
  {\doibase 10.1086/309179} {\bibfield  {journal} {\bibinfo  {journal}
  {Astrophys. J.}\ }\textbf {\bibinfo {volume} {538}},\ \bibinfo {pages} {473}
  (\bibinfo {year} {2000})},\ \Eprint {http://arxiv.org/abs/astro-ph/9911177}
  {arXiv:astro-ph/9911177 [astro-ph]} \BibitemShut {NoStop}%
\bibitem [{\citenamefont {Ade}\ \emph {et~al.}(2016{\natexlab{b}})\citenamefont
  {Ade} \emph {et~al.}}]{Ade:2015xua}%
  \BibitemOpen
  \bibfield  {author} {\bibinfo {author} {\bibfnamefont {P.~A.~R.}\
  \bibnamefont {Ade}} \emph {et~al.} (\bibinfo {collaboration} {Planck}),\
  }\href {\doibase 10.1051/0004-6361/201525830} {\bibfield  {journal} {\bibinfo
   {journal} {Astron. Astrophys.}\ }\textbf {\bibinfo {volume} {594}},\
  \bibinfo {pages} {A13} (\bibinfo {year} {2016}{\natexlab{b}})},\ \Eprint
  {http://arxiv.org/abs/1502.01589} {arXiv:1502.01589 [astro-ph.CO]}
  \BibitemShut {NoStop}%
\bibitem [{\citenamefont {Aiola}\ \emph {et~al.}(2015)\citenamefont {Aiola},
  \citenamefont {Wang}, \citenamefont {Kosowsky}, \citenamefont
  {Kahniashvili},\ and\ \citenamefont {Firouzjahi}}]{Aiola:2015rqa}%
  \BibitemOpen
  \bibfield  {author} {\bibinfo {author} {\bibfnamefont {S.}~\bibnamefont
  {Aiola}}, \bibinfo {author} {\bibfnamefont {B.}~\bibnamefont {Wang}},
  \bibinfo {author} {\bibfnamefont {A.}~\bibnamefont {Kosowsky}}, \bibinfo
  {author} {\bibfnamefont {T.}~\bibnamefont {Kahniashvili}}, \ and\ \bibinfo
  {author} {\bibfnamefont {H.}~\bibnamefont {Firouzjahi}},\ }\href {\doibase
  10.1103/PhysRevD.92.063008} {\bibfield  {journal} {\bibinfo  {journal} {Phys.
  Rev.}\ }\textbf {\bibinfo {volume} {D92}},\ \bibinfo {pages} {063008}
  (\bibinfo {year} {2015})},\ \Eprint {http://arxiv.org/abs/1506.04405}
  {arXiv:1506.04405 [astro-ph.CO]} \BibitemShut {NoStop}%
\bibitem [{\citenamefont {Byrnes}\ \emph
  {et~al.}(2016{\natexlab{b}})\citenamefont {Byrnes}, \citenamefont {Regan},
  \citenamefont {Seery},\ and\ \citenamefont {Tarrant}}]{Byrnes:2015dub}%
  \BibitemOpen
  \bibfield  {author} {\bibinfo {author} {\bibfnamefont {C.~T.}\ \bibnamefont
  {Byrnes}}, \bibinfo {author} {\bibfnamefont {D.}~\bibnamefont {Regan}},
  \bibinfo {author} {\bibfnamefont {D.}~\bibnamefont {Seery}}, \ and\ \bibinfo
  {author} {\bibfnamefont {E.~R.~M.}\ \bibnamefont {Tarrant}},\ }\href
  {\doibase 10.1088/1475-7516/2016/06/025} {\bibfield  {journal} {\bibinfo
  {journal} {JCAP}\ }\textbf {\bibinfo {volume} {1606}},\ \bibinfo {pages}
  {025} (\bibinfo {year} {2016}{\natexlab{b}})},\ \Eprint
  {http://arxiv.org/abs/1511.03129} {arXiv:1511.03129 [astro-ph.CO]}
  \BibitemShut {NoStop}%
\bibitem [{\citenamefont {Quartin}\ and\ \citenamefont
  {Notari}(2015)}]{Quartin:2014yaa}%
  \BibitemOpen
  \bibfield  {author} {\bibinfo {author} {\bibfnamefont {M.}~\bibnamefont
  {Quartin}}\ and\ \bibinfo {author} {\bibfnamefont {A.}~\bibnamefont
  {Notari}},\ }\href {\doibase 10.1088/1475-7516/2015/01/008} {\bibfield
  {journal} {\bibinfo  {journal} {JCAP}\ }\textbf {\bibinfo {volume} {1501}},\
  \bibinfo {pages} {008} (\bibinfo {year} {2015})},\ \Eprint
  {http://arxiv.org/abs/1408.5792} {arXiv:1408.5792 [astro-ph.CO]} \BibitemShut
  {NoStop}%
\bibitem [{\citenamefont {Kogo}\ and\ \citenamefont
  {Komatsu}(2006)}]{Kogo:2006kh}%
  \BibitemOpen
  \bibfield  {author} {\bibinfo {author} {\bibfnamefont {N.}~\bibnamefont
  {Kogo}}\ and\ \bibinfo {author} {\bibfnamefont {E.}~\bibnamefont {Komatsu}},\
  }\href {\doibase 10.1103/PhysRevD.73.083007} {\bibfield  {journal} {\bibinfo
  {journal} {Phys. Rev.}\ }\textbf {\bibinfo {volume} {D73}},\ \bibinfo {pages}
  {083007} (\bibinfo {year} {2006})},\ \Eprint
  {http://arxiv.org/abs/astro-ph/0602099} {arXiv:astro-ph/0602099 [astro-ph]}
  \BibitemShut {NoStop}%
\bibitem [{\citenamefont {Bramante}\ \emph {et~al.}(2013)\citenamefont
  {Bramante}, \citenamefont {Kumar}, \citenamefont {Nelson},\ and\
  \citenamefont {Shandera}}]{Bramante:2013moa}%
  \BibitemOpen
  \bibfield  {author} {\bibinfo {author} {\bibfnamefont {J.}~\bibnamefont
  {Bramante}}, \bibinfo {author} {\bibfnamefont {J.}~\bibnamefont {Kumar}},
  \bibinfo {author} {\bibfnamefont {E.}~\bibnamefont {Nelson}}, \ and\ \bibinfo
  {author} {\bibfnamefont {S.}~\bibnamefont {Shandera}},\ }\href {\doibase
  10.1088/1475-7516/2013/11/021} {\bibfield  {journal} {\bibinfo  {journal}
  {JCAP}\ }\textbf {\bibinfo {volume} {1311}},\ \bibinfo {pages} {021}
  (\bibinfo {year} {2013})},\ \Eprint {http://arxiv.org/abs/1307.5083}
  {arXiv:1307.5083 [astro-ph.CO]} \BibitemShut {NoStop}%
\bibitem [{\citenamefont {Ade}\ \emph {et~al.}(2014{\natexlab{b}})\citenamefont
  {Ade} \emph {et~al.}}]{Ade:2013kta}%
  \BibitemOpen
  \bibfield  {author} {\bibinfo {author} {\bibfnamefont {P.~A.~R.}\
  \bibnamefont {Ade}} \emph {et~al.} (\bibinfo {collaboration} {Planck}),\
  }\href {\doibase 10.1051/0004-6361/201321573} {\bibfield  {journal} {\bibinfo
   {journal} {Astron. Astrophys.}\ }\textbf {\bibinfo {volume} {571}},\
  \bibinfo {pages} {A15} (\bibinfo {year} {2014}{\natexlab{b}})},\ \Eprint
  {http://arxiv.org/abs/1303.5075} {arXiv:1303.5075 [astro-ph.CO]} \BibitemShut
  {NoStop}%
\bibitem [{\citenamefont {Aylor}\ \emph {et~al.}(2017)\citenamefont {Aylor}
  \emph {et~al.}}]{Aylor:2017haa}%
  \BibitemOpen
  \bibfield  {author} {\bibinfo {author} {\bibfnamefont {K.}~\bibnamefont
  {Aylor}} \emph {et~al.} (\bibinfo {collaboration} {SPT}),\ }\href {\doibase
  10.3847/1538-4357/aa947b} {\bibfield  {journal} {\bibinfo  {journal}
  {Astrophys. J.}\ }\textbf {\bibinfo {volume} {850}},\ \bibinfo {pages} {101}
  (\bibinfo {year} {2017})},\ \Eprint {http://arxiv.org/abs/1706.10286}
  {arXiv:1706.10286 [astro-ph.CO]} \BibitemShut {NoStop}%
\bibitem [{\citenamefont {Pearson}\ \emph {et~al.}(2012)\citenamefont
  {Pearson}, \citenamefont {Lewis},\ and\ \citenamefont
  {Regan}}]{Pearson:2012ba}%
  \BibitemOpen
  \bibfield  {author} {\bibinfo {author} {\bibfnamefont {R.}~\bibnamefont
  {Pearson}}, \bibinfo {author} {\bibfnamefont {A.}~\bibnamefont {Lewis}}, \
  and\ \bibinfo {author} {\bibfnamefont {D.}~\bibnamefont {Regan}},\ }\href
  {\doibase 10.1088/1475-7516/2012/03/011} {\bibfield  {journal} {\bibinfo
  {journal} {JCAP}\ }\textbf {\bibinfo {volume} {1203}},\ \bibinfo {pages}
  {011} (\bibinfo {year} {2012})},\ \Eprint {http://arxiv.org/abs/1201.1010}
  {arXiv:1201.1010 [astro-ph.CO]} \BibitemShut {NoStop}%
\end{thebibliography}%

\appendix
\section*{Appendix: Evaluation of the trispectrum in the collapsed limit}
\label{app:trispec_approx}
Following the approximation in \cite{Pearson:2012ba} for the $n=0$ case, we can approximate the integral in Eq.(\ref{eq:reducedtrispectrum}) as products of separate integrals over $r$. For $L \ll \ell_1, \ell_2, \ell_3, \ell_4$, $\alpha_\ell(r)$'s are sharply peaked around $r=r_{\rm cmb}$ and $F_L(r_1, r_2) \approx F_L(r_{\rm cmb}, r_{\rm cmb})$ varies slowly for $r$ values where the other terms are contributing. Then,

\begin{align}
\mathcal{T}^{w\ell_1 x\ell_2}_{y\ell_3 z\ell_4}(L) \approx~ &\taunl h_{\ell_1 
\ell_2 L} h_{\ell_3 \ell_4 L} F_L(r_{\rm cmb}) \nonumber \\ &~~ D^{wx}(\ell_1, 
\ell_2, n) D^{yz}(\ell_3, \ell_4, n)
\end{align}
where,
\begin{align}
    D^{wx}(\ell_1, \ell_2, n) = \int dr\; r^2 \alpha_{\ell_1}^w(r,n) 
    \beta_{\ell_2}^x(r)
    \label{eq:Al1l2n}
\end{align}

We have tested that when the smallest multipole used is $\ell=30$, the approximation provides results within $2.4\%$ percent, and quickly improves to sub-percent level accuracy for $\ell \approx 100$. This allows for fast evaluation of non-Gaussian covariance matrices for dipole modulation parameters. Further, with the following ansatz:

\begin{align}
    D_{L=1}^{wx}(\ell, \ell+1, n) = B_\ell^{wx}(n) 
    \left(\frac{\ell}{\ell_0}\right)^n \sqrt{C_\ell^{wx} C_{\ell+1}^{wx}}
\end{align}
we can interpolate $B_\ell(n)$ using the exact integral values of Eq.(\ref{eq:Al1l2n}) on a $\ell, n$ grid, and use it for a likelihood or MCMC analysis to fit for both $\taunl$ and $n$.

\end{document}